\documentclass[twocolumn]{aastex6}

\usepackage{amsmath}

\def\roma{1}
\def\pescara{2}
\def\nice{3}
\def\rio{4}
\def\cbpf{5}
\def\campus{6}
\def\icracampus{7}

\begin{document}

\title{On the rate and on the gravitational wave emission of short and long GRBs}

\author{R.~Ruffini\altaffilmark{\roma,\pescara,\nice,\rio},
				J.~Rodriguez\altaffilmark{\roma,\pescara},
				M.~Muccino\altaffilmark{\roma,\pescara},
				J.~A.~Rueda\altaffilmark{\roma,\pescara,\rio},
				Y.~Aimuratov\altaffilmark{\roma,\pescara},
				U.~Barres~de~Almeida\altaffilmark{\rio,\cbpf},
				L.~Becerra\altaffilmark{\roma,\pescara},
				C.~L.~Bianco\altaffilmark{\roma,\pescara},
				C.~Cherubini\altaffilmark{\campus,\icracampus},
				S.~Filippi\altaffilmark{\campus,\icracampus},
				D.~Gizzi\altaffilmark{\roma},
				M.~Kovacevic\altaffilmark{\roma,\pescara,\nice},
				R.~Moradi\altaffilmark{\roma,\pescara},
				F.~G.~Oliveira\altaffilmark{\roma,\pescara,\nice},
				G.~B.~Pisani\altaffilmark{\roma,\pescara}, 
				Y.~Wang\altaffilmark{\roma,\pescara}
				}
				
\altaffiltext{\roma}{Dipartimento di Fisica and ICRA, 
                     Sapienza Universit\`a di Roma, 
                     P.le Aldo Moro 5, 
                     I--00185 Rome, 
                     Italy}
                     
\altaffiltext{\pescara}{ICRANet, 
                     P.zza della Repubblica 10, 
                     I--65122 Pescara, 
                     Italy}

\altaffiltext{\nice}{Universit\'e de Nice Sophia Antipolis, 
                     CEDEX 2, Grand Ch\^{a}teau Parc Valrose, 
                     Nice, 
                     France}

\altaffiltext{\rio}{ICRANet-Rio, 
                     Centro Brasileiro de Pesquisas F\'isicas, 
                     Rua Dr. Xavier Sigaud 150, 
                     22290--180 Rio de Janeiro, 
                     Brazil}

\altaffiltext{\cbpf}{Centro Brasileiro de Pesquisas F\'isicas, 
                     Rua Dr. Xavier Sigaud 150, 
                     22290--180 Rio de Janeiro, 
                     Brazil}

\altaffiltext{\campus}{Unit of Nonlinear Physics and Mathematical Modeling, 
                     Universit\`a Campus Bio-Medico di Roma, 
                     Via A.~del Portillo 21, 
                     I--00128 Rome, 
                     Italy}

\altaffiltext{\icracampus}{ICRA, 
                     Universit\`a Campus Bio-Medico di Roma, 
                     Via A.~del Portillo 21, 
                     I--00128 Rome, 
                     Italy}
							
\date{\today}

\begin{abstract}
On the ground of the large number of gamma-ray bursts (GRBs) detected with cosmological redshift, we have introduced a new classification of GRBs in seven subclasses, {all with binary progenitors originating gravitational waves (GWs)}. Each binary is composed by a different combination of carbon-oxygen cores (CO$_{\rm core}$), neutron stars (NSs), black holes (BHs) and white dwarfs (WDs). {This opens an ample new scenario for the role of GWs both as detectable sources and as a determining factor in the coalescence process of the GRB binary progenitors}. The long bursts, traditionally assumed to originate from a single BH with an ultra-relativistic jetted emission, not expected to emit GWs, have instead been subclassified as (I) X-ray flashes (XRFs), (II) binary-driven hypernovae (BdHNe), and (III) BH-supernovae (BH-SNe). They are framed within the induced gravitational collapse (IGC) paradigm {with progenitor a tight binary composed of a CO$_{\rm core}$ and a NS or BH companion. The supernova (SN) explosion of the CO$_{\rm core}$ triggers a hypercritical accretion process onto the companion NS or BH. If the accretion is not sufficient for the NS to reach its critical mass, an XRF occurs, while when the BH is already present or formed by the hypercritical accretion, a BdHN occurs. In the case these binaries are not disrupted by the mass-loss process, XRFs lead to NS-NS binaries and BdHNe lead to NS-BH ones}. The short bursts, originating in NS-NS mergers, are subclassified as (IV) short gamma-ray flashes (S-GRFs) and (V) short GRBs (S-GRBs), the latter when a BH is formed. Two additional families are (VI) ultra-short GRBs (U-GRBs) and (VII) gamma-ray flashes (GRFs), respectively formed in NS-BH and NS-WD mergers. We use the estimated occurrence rate of the above subclasses and their GW emission to assess their detectability by Advanced LIGO and Virgo, eLISA, and resonant bars. {We also discuss the consequences of our results in view of the recent announcement of the LIGO-Virgo Collaboration of the source GW 170817 as being originated by a NS-NS merger.}
\end{abstract}


\maketitle

\section{Introduction}\label{sec:1}

Thanks to the extensive observations carried out by $\gamma$-ray telescopes, such as AGILE, BATSE, BeppoSAX, \emph{Fermi}, \emph{HETE-II}, \emph{INTEGRAL}, Konus/WIND and \emph{Swift}, our understanding of ``long'' and ``short'' gamma-ray burst (GRB) progenitor systems has greatly improved. This has led also to a vast literature devoted to the estimate of their relative occurrence rates, all in general agreement. For long bursts see,	 e.g.,~\citet{Soderberg2006Nature,2007ApJ...657L..73G,2007ApJ...662.1111L,2009MNRAS.392...91V,2010tsra.confE.204R,2010MNRAS.406.1944W,2011A&A...525A..53G,Kovacevic2014}; for short bursts see, e.g., \citet{Virgili2011,2015MNRAS.448.3026W}; and for both long and short bursts see, e.g., \citet{2015ApJ...812...33S,2016ApJ...832..136R}.	
The rates of GW emission from GRBs have been calculated in the literature at a time in which short GRBs were considered to be originated in neutron star-neutron star (NS-NS) binaries, while long GRBs, were considered to be originated in single events\footnote{{With the exception of the binary progenitors proposed in \citet{1998ApJ...502L...9F,1999ApJ...526..152F,1999ApJ...520..650F,2002ApJ...571..394B}.}}, e.g.~\emph{collapsars} (\citealp{1993ApJ...405..273W,1999ApJ...524..262M,2001ApJ...550..410M,2006ARA&A..44..507W}; {see however \citealp{2018ApJ...852...53R}) and \emph{magnetars} (\citealp{1992Natur.357..472U,1998A&A...333L..87D,1998PhRvL..81.4301D,1998ApJ...505L.113K,2001ApJ...552L..35Z}; see however \citealp{2016ApJ...832..136R}).} Thus, only short GRBs have been up to now considered to estimate the simultaneous detection rate of gravitational waves (GWs) and GRBs. For instance, \citet{2015MNRAS.448.3026W} used the luminosity function of short GRBs observed by Swift; \citet{2014ApJ...789...65Y} by BATSE; \citet{2016JCAP...11..056P} by Fermi and \citet{2016A&A...594A..84G} by Swift and Fermi. 

In our recent works \citep[see][and references therein]{2016ApJ...832..136R} we have introduced a new {classification} in which all GRBs, namely both long and short, originate from merging and/or accreting binary systems, each composed by a different combination of carbon-oxygen cores (CO$_{\rm core}$), NSs, black holes (BHs) and white dwarfs (WDs). For each system the initial state and the final state are respectively here referred to as ``\emph{in-state}'' and ``\emph{out-state}''. {This opens an ample new scenario for the role of GWs both as detectable sources and as a determining factor in the coalescence process of the GRB progenitors.}

We interpret the traditional long GRBs within the induced gravitational collapse (IGC) paradigm \citep{2006tmgm.meet..369R,Ruffini2007b,2008mgm..conf..368R,2012A&A...548L...5I,2012ApJ...758L...7R,2014ApJ...793L..36F,2015ApJ...798...10R} that proposes as \emph{in-state} a tight binary system composed of a CO$_{\rm core}$ undergoing a supernova (SN) explosion and a companion compact object, e.g. a NS (or a BH). The SN explosion triggers a hypercritical accretion onto the NS companion, {whose details has been studied, simulated and presented in several publications \citep[see, e.g.,][references therein and Appendix~\ref{app:A}]{2016ApJ...833..107B,2015PhRvL.115w1102F,2015ApJ...812..100B,2014ApJ...793L..36F}. Depending upon the binary parameters, the hypercritical accretion can lead to three very different outcomes:}

\begin{table*}
\caption{\label{tab:rates} Summary of the astrophysical aspects of the different GRB subclasses and of their observational properties. In the first four columns we indicate the GRB subclasses and their corresponding \emph{in-states} and the \emph{out-states}. In columns 5--8 we list the ranges of $E_{\rm p,i}$ and $E_{\rm iso}$ (rest-frame $1$--$10^4$~keV), $E_{\rm iso,X}$ (rest-frame $0.3$--$10$~keV), and $E_{\rm iso,GeV}$ (rest-frame $0.1$--$100$~GeV). Columns 9 and 10 list, for each GRB subclass, the maximum observed redshift and the local observed rate $\rho_{\rm GRB}$ obtained in \citet{2016ApJ...832..136R}. {We refer the reader to Appendix~\ref{app:B} for details on the method used to calculate $\rho_{\rm GRB}$.}}
\begin{ruledtabular}
\begin{tabular}{llcccccccc}
     & subclass  & \emph{In-state}  & \emph{Out-state} & $E_{\rm p,i}$ &  $E_{\rm iso}$ & $E_{\rm iso,X}$ &  $E_{\rm iso,Gev}$ &  $z_{\rm max}$ & $\rho_{\rm GRB}$ \\
& & & & (MeV) & (erg) & (erg) & (erg) & & (Gpc$^{-3}$yr$^{-1}$)\\  		
\hline
I    & XRFs & CO$_{\rm core}$-NS    & $\nu$NS-NS & $\lesssim0.2$  &  $\sim 10^{48}$--$10^{52}$ &  $\sim 10^{48}$--$10^{51}$ &  $-$ &  $1.096$ & $100^{+45}_{-34}$\\
II   & BdHNe  & CO$_{\rm core}$-NS  & $\nu$NS-BH & $\sim0.2$--$2$ &  $\sim 10^{52}$--$10^{54}$ &  $\sim 10^{51}$--$10^{52}$ &  $\lesssim 10^{53}$ &  $9.3$ & $0.77^{+0.09}_{-0.08}$\\
III  & BH-SN & CO$_{\rm core}$-BH  & $\nu$NS-BH & $\gtrsim2$ &  $>10^{54}$ & $\sim 10^{51}$--$10^{52}$ &  $\gtrsim 10^{53}$ &  $9.3$ & $\lesssim 0.77^{+0.09}_{-0.08}$\\
IV   & S-GRFs & NS-NS & MNS & $\lesssim2$ &  $\sim 10^{49}$--$10^{52}$ &  $\sim 10^{49}$--$10^{51}$ &  $-$ &  $2.609$ &  $3.6^{+1.4}_{-1.0}$\\
V    & S-GRBs  & NS-NS & BH & $\gtrsim2$ &  $\sim 10^{52}$--$10^{53}$ &  $\lesssim 10^{51}$ &  $\sim 10^{52}$--$10^{53}$ &  $5.52$  & $\left(1.9^{+1.8}_{-1.1}\right)\times10^{-3}$\\
VI   & U-GRBs & $\nu$NS-BH & BH & $\gtrsim2$ &  $>10^{52}$ &  $-$ & $-$ &  $-$ & $\gtrsim 0.77^{+0.09}_{-0.08}$\\
VII  & GRFs  & NS-WD & MNS & $\sim0.2$--$2$ &  $\sim 10^{51}$--$10^{52}$ &  $\sim 10^{49}$--$10^{50}$ & $-$ &  $2.31$ & $1.02^{+0.71}_{-0.46}$\\
\end{tabular}
\end{ruledtabular}
\end{table*}

\begin{itemize}
\item[I.]
X-ray flashes (XRFs) with isotropic energy $E_{\rm iso}\lesssim 10^{52}$~erg and rest-frame spectral peak energy $E_{p,i}\lesssim 200$~keV. 
This class occurs in CO$_{\rm core}$--NS binaries when the hypercritical accretion onto the NS companion is not enough to induce gravitational collapse into a BH \citep{2016ApJ...833..107B,2015ApJ...812..100B}. Following this definition, \citet{2016ApJ...832..136R} estimated for the XRF a local observed rate of $\rho_{\rm XRF}=100^{+45}_{-34}$~Gpc$^{-3}$~yr$^{-1}$ \citep{2016ApJ...832..136R}. This rate is in agreement with that of low-luminous long GRBs{, e.g.:$325^{+352}_{-177}$~Gpc$^{-3}$~yr$^{-1}$ \citep{2007ApJ...662.1111L}, $\sim 200$~Gpc$^{-3}$~yr$^{-1}$\citep{2009MNRAS.392...91V}, $164^{+98}_{-65}$~Gpc$^{-3}$~yr$^{-1}$ \citep{2015ApJ...812...33S}}.
After the SN explosion the binary can either get disrupted or remain bound depending upon the mass loss and/or natal kick imparted to the system \citep[see][references therein {and Appendix~\ref{app:A5}}]{2014LRR....17....3P}. In the former case the XRF leads to two runaway NSs, while in the latter one, the \emph{out-states} of XRFs are binaries composed of a newly-formed $\sim 1.4$--$1.5~M_\odot$ NS (hereafter $\nu$NS) born in the SN explosion, and a massive NS (MNS) which accreted matter from the SN ejecta. Typical periods of these binaries are $P_{\rm orb} \gtrsim 30$~min \citep{2016ApJ...833..107B}.
\item[II.]
Binary-driven hypernovae (BdHNe) with $E_{\rm iso}\gtrsim10^{52}$~erg and $E_{p,i}\gtrsim200$~keV. 
BdHNe occur in more compact CO$_{\rm core}$--NS binaries which leads to a more massive hypercritical accretion onto the NS, hence leading to BH formation. {Following this definition, \citet{2016ApJ...832..136R} estimated for the BdHNe a local observed rate} $\rho_{\rm BdHN}=0.77^{+0.09}_{-0.08}$~Gpc$^{-3}$~yr$^{-1}$ \citep{2016ApJ...832..136R}. This rate is in agreement with that for high-luminous long GRBs{, e.g.: $1.3^{+0.6}_{-0.7}$~Gpc$^{-3}$~yr$^{-1}$ \citep{2010MNRAS.406.1944W} and $0.8^{+0.1}_{-0.1}$~Gpc$^{-3}$~yr$^{-1}$ \citep{2015ApJ...812...33S}}. As in the case of XRFs, the SN explosion can disrupt the binary depending upon the mass loss and/or natal kick. In the case when the system remains bound, the \emph{out-states} of BdHNe are $\nu$NS-BH binaries \citep[see][{and Appendix~\ref{app:A5}}]{2015PhRvL.115w1102F}. Typical periods of these binaries are 5~min$\lesssim P_{\rm orb} \lesssim 30$~min \citep{2016ApJ...833..107B}.

\item[III.]
BH-SN with $E_{\rm iso}\gtrsim10^{54}$~erg and $E_{p,i}\gtrsim2$~MeV. BH-SN occur in close CO$_{\rm core}$ ({or Helium or Wolf-Rayet star})-BH binaries \citep{Ruffini2001c} in which the hypercritical accretion occurs onto a previously formed BH. Such BH-SN systems correspond to the late evolutionary stages of X-ray binaries as Cyg X-1 \citep{1978pans.proc.....G,2011ApJ...742L...2B}, or microquasars \citep{1998Natur.392..673M}. {These systems might be also formed following the binary evolutionary patch leading to the scenario XI in \citet{1999ApJ...526..152F}. Since the estimated rate of BdHNe covers systems with the above $E_{\rm iso}$ and $E_{p,i}$ range, we can adopt} the rate of BdHNe as an upper limit to the rate of BH-SNe, i.e. $\rho_{\rm BH-SN}\lesssim \rho_{\rm BdHN} = 0.77^{+0.09}_{-0.08}$~Gpc$^{-3}$~yr$^{-1}$ \citep{2016ApJ...832..136R}. As in the above cases of XRFs and BdHNe, the SN explosion may disrupt the binary. If the binary survives, then the \emph{out-states} of BH-SNe can be a $\nu$NS-BH or a BH-BH if the SN central remnant directly collapses to a BH. However, the latter scenario is currently ruled out by the observations of pre-SN cores which appear to have masses $\lesssim 18~M_\odot$, very low to lead to direct BH formation \citep[see, e.g.,][for details]{2009ARA&A..47...63S,2015PASA...32...16S}.
\end{itemize}

In the current literature such a difference between an XRF, a BdHN and a BH-SN in the evaluation of GWs, here implemented, is still missing.

We turn now to the short bursts. Although their progenitors are still under debate, there is an ample consensus in the scientific community that they originate from NS-NS and/or NS-BH merging binaries \citep[see, e.g.,][]{Goodman1986,Paczynski1986,Eichler1989,Narayan1991,MeszarosRees1997_b,Rosswog2003,Lee2004,2014ARA&A..52...43B}. By adopting the same \emph{in-states} as in the above traditional models, namely NS-NS and/or NS-BH mergers, they can be divided into three subclasses \citep{2015PhRvL.115w1102F,2015ApJ...808..190R,2016ApJ...832..136R}:
\begin{itemize}
\item[IV.]
Short gamma-ray flashes (S-GRFs), with $E_{\rm iso}\lesssim10^{52}$~erg and $E_{p,i}\lesssim2$~MeV, occur when no BH is formed in the NS-NS merger, i.e. they lead to a MNS. Following this definition, \citet{2016ApJ...832..136R} estimated for the S-GRFs a local observed rate $\rho_{\rm S-GRF}=3.6^{+1.4}_{-1.0}$~Gpc$^{-3}$~yr$^{-1}$.
\item[V.]
Authentic short GRBs (S-GRBs), with $E_{\rm iso}\gtrsim 10^{52}$~erg and $E_{p,i}\gtrsim2$~MeV, occur when a BH is formed in the NS-NS merger \citep{2016ApJ...831..178R,2015ApJ...808..190R,2013ApJ...763..125M}. Following this definition, \citet{2016ApJ...832..136R} estimated for the S-GRBs a local observed rate $\rho_{\rm S-GRB}=\left(1.9^{+1.8}_{-1.1}\right)\times10^{-3}$~Gpc$^{-3}$~yr$^{-1}$ \citep{2016ApJ...832..136R}.
\item[VI.]
Ultra-short GRBs (U-GRBs), a new subclass of short bursts originating from $\nu$NS-BH merging binaries. They can originate from BdHNe (see II above) or from BH-SN events (see III above). Since in \citet{2015PhRvL.115w1102F} it was shown that the majority of BdHN \emph{out-states} remain bound, we can assume as upper limit of their local density rate, $\rho_{\rm U-GRB} \approx \rho_{\rm BdHN} = 0.77^{+0.09}_{-0.08}$~Gpc$^{-3}$~yr$^{-1}$ \citep{2016ApJ...832..136R}. U-GRBs are yet unobserved/unidentified and present a great challenge not only in the case of high-energy but also possibly in the radio band where they could manifest themselves, prior to the merger phase, as pulsar-BH binaries \citep[see, e.g.,][and references therein]{2015aska.confE..39T}.

\end{itemize}

It is important to mention that the sum of the occurrence rates of the above short burst subclasses IV--VI is in agreement with the estimates obtained from the whole short burst population reported in the literature \citep[see, e.g.,][]{2015MNRAS.448.3026W,2015ApJ...812...33S}. It is then clear that what in the current literature are indicated as short GRBs are actually just S-GRFs.

In addition to the above three subclasses of long bursts and three subclasses of short bursts, we recall the existence of a class of bursts occurring in a low-density circumburst medium (CBM), e.g.~$n_{\rm CBM}\sim∼10^{-3}$~cm$^{-3}$, which show hybrid properties between short and long bursts in $\gamma$-rays. These bursts are not associated with SNe, even at low  redshift where the SN detection would not be precluded \citep{DellaValle2006sn}. We have called such bursts as Gamma-ray flashes (GRFs) \citep{2016ApJ...832..136R}.

\begin{itemize}

\item[VII.]
GRFs have $10^{51}\lesssim E_{\rm iso}\lesssim 10^{52}$~erg and $0.2 \lesssim E_{p,i}\lesssim 2$~MeV. These bursts, which shows an extended and softer emission, are thought to originate in NS-WD mergers \citep{2016ApJ...832..136R}. NS-WD binaries are notoriously common astrophysical systems \citep{2015ApJ...812...63C} and possible evolutionary scenarios leading to such mergers have been envisaged \citep[see, e.g.,][]{1999ApJ...520..650F,2014MNRAS.437.1485L,2000ApJ...530L..93T}\footnote{An additional (but less likely) scenario leading to merging NS-WD systems might occur in a NS-NS approaching the merger phase \citep{2016ApJ...832..136R}. According to \citet{1992ApJ...400..175B} and \citet{1977ApJ...215..311C} (see also references therein), in a very close, NS-NS binary with unequal-mass components, stable mass-transfer from the less massive to the more massive NS might occur for appropriate mass-ratios in such a way that the donor NS moves outward in the mass-loss process until it reaches the beta-decay instability becoming a low-mass WD.}. GRFs form a MNS and not a BH \citep[see][for details]{2016ApJ...832..136R}. Following this definition, \citet{2016ApJ...832..136R} estimated for the GRFs a local observed rate $\rho_{\rm GRF}=1.02^{+0.71}_{-0.46}$~Gpc$^{-3}$~yr$^{-1}$ \citep{2016ApJ...832..136R}. {This density rate appears to be low with respect to the current number of known NS-WD binaries in the Galaxy \citep[see, e.g.,][]{2015ApJ...812...63C}. From the GRB side, we note that indeed only one NS-WD merger has been identified \citep[see analysis of GRB 060614 in][]{2009A&A...498..501C}. The above implies that, very likely, the majority of the expected mergers are under the threshold of the existing X and gamma-ray detectors.}
\end{itemize}

The aforementioned density rates for all GRB subclasses have been estimated in \citet{2016ApJ...832..136R} assuming no beaming. {The presence of beaming would require the observation of achromatic jet breaks in the afterglow light curve. In the present case of short bursts such clear achromatic jet breaks have never been observed. \citet{2015ApJ...815..102F} reported 4 measured jet breaks in a sample of 11 short bursts: GRB 051221A, GRB 090426A, GRB 111020A, GRB 130603B (see Table 5 there). However:}

{
- GRB 051221A: The break is inferred only from the X-ray light curve, while the contemporary optical and radio data does not support such an interpretation \citep[see][]{2006ApJ...650..261S}.
}

{
- GRB 090426A: The break is inferred from the optical band only, and there are no contemporary observations in other bands \citep[see][]{2011A&A...531L...6N}.
}

{
- GRB 111020A: The break is inferred only from the X-ray light curve, but this interpretation is based on a single upper limit by Chandra and no data points \citep[see][]{2012ApJ...756..189F}.
}

{
- GRB 130603B: The break is inferred from the optical band and is compatible with the radio data. However, contemporary X-ray observations are clearly contradicting this interpretation and presents no break at all. In fact, the authors invoke the presence of an extra source to justify what they call ``late time X-ray excess'' \citep[see][]{2014ApJ...780..118F}.
}

{
In addition, \citet{2017ApJ...844...83A,2018arXiv180207552R} have shown that, in all the identified S-GRBs, the GeV emission has been always observed when the source was within the Fermi-LAT field of view. This result points as well to no significant presence of beaming in the GeV emission of S-GRBs.
}

{
Therefore, all the above points imply that there is still no evidence for the need to assume beaming.
}

We show in Table~\ref{tab:rates} a summary of the astrophysical aspects related to the GRB subclasses and their observational properties. 

The aim of this article is to use the rate of occurrence of the above GRB subclasses to assess the detectability of their associated GW emission by the ground-based interferometers Advanced LIGO and Advanced Virgo, by the space-based interferometer eLISA, as well as by the resonant bars, for completeness.

We show in Table~\ref{tab:acronyms} a summary of acronyms used in this work.
\begin{table}
\centering
\begin{tabular}{lc}
\hline\hline
Extended wording & Acronym \\
\hline
Binary-driven hypernova & BdHN \\
Black hole                    & BH \\
Carbon-oxygen core      & CO$_{\rm core}$ \\ 
Gamma-ray burst         & GRB \\
Gamma-ray flash          & GRF \\
Induced gravitational collapse & IGC \\
Massive neutron star     & MNS \\
Neutron star                & NS \\
New neutron star created in the SN explosion          & $\nu$NS \\
Short gamma-ray burst  & S-GRB \\
Short gamma-ray flash  & S-GRF \\
Supernova                  & SN \\
Ultrashort gamma-ray burst & U-GRB \\ 
White dwarf                & WD \\
X-ray flash                  & XRF \\
\hline
\end{tabular}
\caption{Acronyms used in this work in alphabetic order.}
\label{tab:acronyms}
\end{table}

\section{Relevance of the NS structure and critical mass}\label{sec:2}

Having introduced the above seven subclasses of GRBs, it becomes clear the relevance of the NS physics, in particular the NS critical mass value, in the definition of the subclasses I-II and IV-V.

First, we recall that in our previous works we have adopted a NS critical mass within the range $2.2$--$3.4~M_\odot$, depending on the equation of state (EOS) and on the NS angular momentum \citep{2015PhRvD..92b3007C,2015ApJ...812..100B,2014NuPhA.921...33B}. These quoted values are for EOS based on relativistic nuclear mean-field models (in this case the NL3, TM1 and GM1 models) and for a NS angular momentum from $J=0$ up to $J_{\rm max}\approx 0.7 G M^2/c$ \citep{2015PhRvD..92b3007C}. Hereafter, we adopt the stiffest model, namely the NL3 EOS, which leads to the largest NS critical mass: from $M_{\rm crit}\approx 2.7~M_\odot$ at $J=0$, that, as expected, is lower than the non-rotating critical mass upper limit of $3.2~M_\odot$ established by \citet{1974PhRvL..32..324R}, to $M_{\rm crit}\approx 3.4~M_\odot$ at $J_{\rm max}$ \citep{2015PhRvD..92b3007C}. Our choice of relativistic mean-field theory models is based on the fact that they satisfy important properties such as Lorentz covariance, relativistic self-consistency (hence they do not violate causality), intrinsic inclusion of spin, and a simple mechanism of nuclear matter saturation \citep[see, e.g.,][for further details on these kind of models]{PhysRevC.90.055203,PhysRevC.93.025806}. The above three representative EOS that we have explored satisfy in addition the astrophysical constraint of leading to a NS critical mass larger than the heaviest massive NS observed, PSR J0348+0432, with $M = 2.01 \pm 0.04 M_\odot$ \citep{2013Sci...340..448A}.

As discussed in \citet{2016ApJ...832..136R}, the separatrix energy value of $\approx 10^{52}$~erg between the subclasses I and II appears as a theoretical estimate of the upper limit to the energy emitted in the hypercritical accretion process onto a $\sim 1.4~M_\odot$ NS \citep[see, e.g.,][]{2016ApJ...833..107B} and the afore-mentioned adopted critical mass. This has been shown to be in agreement with the observations of 20 XRFs and 233 BdHNe (up to the end of 2014). In fact, observationally, the current upper limit for XRFs is $(7.3\pm 0.7)\times 10^{51}$~erg, and the lower limit for BdHNe is $(9.2\pm 1.3)\times 10^{51}$~erg \citep[see][for further details]{2016ApJ...832..136R}. It is clear that the separatrix energy should have some dependence on the initial NS mass undergoing accretion and on the precise value of the non-rotating critical mass. Although the precise value of the latter is yet unknown, it is constrained within the range $2.0$--$3.2~M_\odot$, where the lower value is the mass of PSR J0348+0432, and the upper value is the well-established absolute maximum NS mass of \citet{1974PhRvL..32..324R}.

It is clear that similar arguments apply also to the case of the subclasses IV and V \citep{2015ApJ...808..190R}; namely the amount of energy emitted during the NS-NS merger leading to a BH should be $\gtrsim 10^{52}$~erg. Observationally, the current upper limit for S-GRFs is $(7.8\pm 1.0)\times 10^{51}$~erg, and the lower limit for BdHNe is $(2.44\pm 0.22)\times 10^{52}$~erg \citep[see][for further details]{2016ApJ...832..136R}.

The above sub-classification is further supported by the fact that GeV emission, expected in presence of a rotating BH, is indeed observed only in BdHNe \citep[e.g.][]{2015ApJ...798...10R} and in S-GRBs \citep[e.g.][]{2017arXiv170408179A,2016ApJ...831..178R,2015ApJ...808..190R,2013ApJ...763..125M}, and absent in XRFs and S-GRFs where no BH is formed \citep[see Figure 10 and the Appendix in][for more details]{2016ApJ...832..136R}.

Therefore, the direct observation of the separatrix energy between XRFs and BdHNe, as well as between S-GRFs and S-GRBs, and their precise occurrence rates ratio, give crucial information on the actual NS critical mass value.

\section{Ingredients Set-up for the Computation of the GW Emission and its Detectability}\label{sec:3}

We have recalled in section~\ref{sec:1} that the evolution of the binary progenitors of both short and long GRBs lead to compact binaries which will eventually merge in a characteristic timescale and emit GWs. We turn now in the following sections to assess the detectability of the GW emission by these merging binaries by Advanced LIGO. 

In order to do this, we make the following drastic simplified assumptions:
\begin{enumerate}
\item
Although it is manifest that the release of gravitational energy of the system in the merger phase is dominated by the X, gamma-ray and GeV emission (see Table~\ref{tab:rates}), we assume that the binary dynamics is only driven by the GW emission.
\item
Consistent with the above GW emission dominance assumption, we further assume that the GW waveform is known and thus one can use the matched filtering technique to estimate the signal-to-noise ratio. The actual GW waveform under the realistic conditions of electromagnetic emission dominance is still unknown. 
\item
To estimate the maximum distance of GW detectability we adopt optimally oriented sources with respect to the detector.
\end{enumerate}
The above assumptions are made with the only aim of establishing an absolute upper limit to the GW emission and its putative detectability under the most optimistic conditions. Similarly, we assume that the binarity of the system does not compromise the interior structure of the NS (see Sec.~\ref{sec:2}).

The minimum GW frequency detectable by the broadband Advanced LIGO interferometer is $f^{\rm aLIGO}_{\rm min}\approx 10$~Hz \citep{2015CQGra..32g4001L}. Since during the binary inspiral the GW frequency is twice the orbital one, the above implies a binary is inside the Advanced LIGO band for orbital periods $P_{\rm orb}\lesssim 0.2$~s. 

\subsection{Systems to be analyzed}\label{sec:3a}

The CO$_{\rm core}$-NS binaries, \emph{in-states} of XRFs and BdHNe, and CO$_{\rm core}$-BH binaries, \emph{in-states} of BH-SN, are not detectable by Advanced LIGO since they have orbital periods $P_{\rm orb} \gtrsim 5$~min~$\gg 0.2$~s \citep{2016ApJ...833..107B}. {After their corresponding hypercritical accretion processes, it is clear that the \emph{out-states} of both XRFs and BdHNe can become the \emph{in-states} of short GRBs, as follows \citep{2016ApJ...832..136R,2015PhRvL.115w1102F,2015ApJ...812..100B}}. 

First let us discuss the \emph{out-states} of XRFs. We have mentioned that XRFs can either get disrupted by the SN and lead to runaway NSs or, in the case the binary remains bound, lead to a $\nu$NS-NS system. Since $\rho_{\rm XRF}>\rho_{\rm S-GRF}+\rho_{\rm S-GRB}$, such $\nu$NS-NS binaries, \emph{out-states} of XRFs, could be the \emph{in-states} of S-GRFs (NS-NS mergers leading to a MNS) and/or S-GRBs (NS-NS mergers leading to a BH). By denoting the total rate of short bursts as $\rho_{\rm short} \equiv \rho_{\rm S-GRF}+\rho_{\rm S-GRB}$, our estimated rates would imply that the fraction of systems which appear to remain bound as $\nu$NS-NS is $(\rho_{\rm short}/\rho_{\rm XRF})\approx 2$\%--$8$\%, while 92\%--98\% of XRFs are disrupted by the SN explosion. Interestingly, this is consistent with the fraction of bound NS-NS obtained in population synthesis analyses \citep[see, e.g.,][references therein {and Appendix~\ref{app:A4} and \ref{app:A5}}]{2012ApJ...759...52D,2014LRR....17....3P,2015ApJ...806..263D,2015ApJ...812...24F,2016ApJ...819..108B}. Therefore, these merging $\nu$NS-NS binaries are clearly included in the S-GRF and S-GRB population. Such binaries are at birth undetectable by Advanced LIGO since they have initially $P_{\rm orb} \gtrsim 5$~min~$\gg 0.2$~s, but their merging can become detectable. 

We have already recalled in the Introduction that in \citet{2015PhRvL.115w1102F} it was shown that, contrary to the case of XRFs, most BdHNe are expected to keep bound after the SN explosion in view of their short orbital periods and more massive accretion process. We have argued that those mergers would lead to the new class of short bursts, the U-GRBs \citep{2015PhRvL.115w1102F}, which however have still to be electromagnetically identified. The same applies to the $\nu$NS-BH systems produced by BH-SN systems, with the only difference being the mass of the BH which, by definition of this subclass, can be larger than the NS critical mass since this BH is formed from direct collapse of a massive star. All the above merging $\nu$NS-BH binaries are, by definition, the U-GRB population. Such binaries are at birth undetectable by Advanced LIGO because their initial orbital periods $P_{\rm orb} \gtrsim 5$~min~$\gg 0.2$~s, but their merger can become detectable.

In the case of NS-WD binaries, the WD large radius and its very likely tidal disruption by the NS make their GW emission hard to be detected \citep[see, e.g.,][]{2009PhRvD..80b4006P}. Thus, we do not consider NS-WD binaries in the following GW discussion.

To summarize, we are going to analyze below the GW emission and detectability of S-GRF and S-GRB, the mergers of $\nu$NS-NS produced by XRFs, as well as of U-GRBs, which are the merger of the $\nu$NS-BH produced by BdHNe and BH-SNe.

\subsection{Binary component masses}\label{sec:3b}

For S-GRFs, we consider the simple case of non-spinning, equal-mass NS-NS merging binaries, i.e. $m_1=m_2=m$. The precise value of the merging NS masses leading to a BH is still poorly known, thus we have chosen as an upper limit roughly half the maximum NS critical mass {(see Sec.~\ref{sec:2})}. Thus, we shall explore mass values $m \approx (1$--$1.7)~M_\odot$.

{For S-GRBs, we also consider non-spinning, equal-mass NS-NS merging binaries. For self-consistency, we choose a range of component masses starting from the upper edge of the S-GRF one}, i.e. $m \approx 1.7~M_\odot$, up to the maximum non-rotating stable mass, i.e.~ $m\approx 2.8~M_\odot$.

For U-GRBs, we adopt in the case of \emph{out-states} of BdHNe, $m_1=1.5~M_\odot$ for the $\nu$NS and $m_{\rm BH} = 2.7$--$3.4~M_\odot$ for the BH {(see Sec.~\ref{sec:2})}. {In the case of \emph{out-states} of BH-SNe, we adopt $m_1=1.5~M_\odot$ for the $\nu$NS and $m_{\rm BH} = 3.4$--$10~M_\odot$ for the BH consistent with the assumption that the BH in this subclass has been previously formed in the binary evolution and therefore it can have a mass larger than the NS critical mass.}

\subsection{Signal-to-noise ratio}\label{sec:3c}

We first recall the main ingredients needed to estimate the detectability of the aforementioned merging binaries associated with the different GRB classes. The signal $h(t)$ induced in the detector is:
\begin{equation}
    h(t) = F_{+}( \theta, \phi, \psi)h_{+}(t, \iota, \beta)+F_{\times}(\theta, \phi, \psi)h_{\times}(t, \iota, \beta),
    \label{eqn:h-induced}
\end{equation}
where $h_{+}$ and $h_{\times}$ are the two polarizations of the GW; $\iota$ and $\beta$ are the polar and azimuthal angles of the unit vector from the source to the detector, relative to a coordinate system centered in the source. The detector pattern functions $F_+$ and $F_{\times}$ depend on the localization of the source with respect to the detector, i.e. they depend on the spherical polar angles $\theta$ and $\phi$ of the source relative to a coordinate system centered in the detector. The pattern functions also depend on the polarization angle $\psi$.  

Since the GW signal might be deep inside the noise, the signal-to-noise ratio, denoted hereafter by $\rho$, is usually computed using the matched filter technique, i.e.~\citep{1998PhRvD..57.4535F}:
\begin{equation}
    \rho^2=4\int_{0}^{\infty} \frac{\bigl \lvert \tilde{h}(f)\bigr \rvert^2}{S_n(f)}df,
    \label{eqn:SNR}
\end{equation}
where $f$ is the GW frequency in the detector's frame, $\tilde{h}(f)$ is the Fourier transform of $h(t)$ and $\sqrt{S_n(f)}$ is the one-sided amplitude spectral density (ASD) of the Advanced LIGO noise. We recall that in the detector's frame the GW frequency is redshifted by a factor $1+z$ with respect to the one in the source's frame, $f_s$, i.e. $f = f_s/(1+z)$. 

The exact position of the binary relative to the detector and the orientation of the binary rotation plane are usually unknown, thus it is a common practice to estimate the signal-to-noise ratio averaging over all the possible locations and orientations, i.e.:
\begin{equation}
    \langle \rho^2 \rangle =  4\int_{0}^{\infty}  \frac{\langle\lvert\tilde{h}(f)\rvert^2\rangle}{S_n(f)}df =4 \int_{0}^{\infty} \frac{h_c^2(f)}{f^2 S_n(f)}df,
    \label{eqn:averaged-SNR-1}
\end{equation}
with $h_c(f)$ the characteristic strain \citep{1998PhRvD..57.4535F}
\begin{equation}
    h_c = \frac{(1+z)}{\pi d_l}\sqrt{\frac{\langle F_{+}^2 \rangle}{2}\frac{G}{c^3}\frac{d E}{df_s}[(1+z) f]},
    \label{eqn:h-char-gen}
\end{equation}
where
\begin{equation}\label{eq:dl}
d_l=\frac{(1+z)c}{H_0} \int_0^z [\Omega_M (1+x)^3+\Omega_\Lambda]^{-1/2} dx,
\end{equation}
is the source luminosity distance and we have used the fact that $\langle F_{+}^2\rangle=\langle F_{\times}^2\rangle$ and $\langle F_{+} F_{\times}\rangle=0$. We recall that $\langle F_{+}^2 \rangle=1/5$ for an interferometer and $\langle F_{+}^2 \rangle=4/15$ for a resonant bar \citep[see, e.g.,][]{maggiore2007gravitational}. We adopt a $\Lambda$CDM cosmology with $H_0 = 71$~km~s$^{-1}$~Mpc$^{-1}$, $\Omega_M=0.27$ and $\Omega_\Lambda = 0.73$ \citep{2015ApJ...802...20R}. It is important to recall that, as we have mentioned, we are interested in estimating the GW detectability under the most optimistic conditions. Thus, to estimate the maximum distance of GW detectability we adopt in Sec.~\ref{sec:3} the ansatz of optimally oriented sources with respect to the detector. The above averaging procedure is here used with the only aim of giving an estimate of the GW strain amplitude, $h_c$, compared and contrasted below in Sec.~\ref{sec:5} with the detectors strain-noise.

\section{GW energy spectrum}\label{sec:4}

In general, a GW-driven binary system evolves in time through two regimes: the first is the \emph{inspiral regime} and the second, which we refer hereafter to as \emph{merger regime}, is composed in the most general case of the final plunge, of the merger, and of the ringdown (oscillations) of the newly formed object. 

\subsection{Inspiral regime}\label{sec:4a}

During the inspiral regime the system evolves describing quasi-circular orbits and it is well described by the traditional point-like quadrupole approximation \citep{1963PhRv..131..435P,1964PhRv..136.1224P,1974bhgw.book.....R,1975ctf..book.....L}. 
The GW frequency is twice the orbital frequency ($f_s= 2 f_{\rm orb}$) and grows monotonically. The energy spectrum during the inspiral regime is:
\begin{equation}\label{eqn:spectrum-inspiral}
    \frac{dE}{df_s} = \frac{1}{3}(\pi G)^{2/3} M_c^{5/3} f_s^{-1/3},
\end{equation}
where $M_c = \mu^{3/5} M^{2/5}=\nu^{3/5} M$ is the called \emph{chirp mass}, $M=m_1+m_2$ is the total binary mass, $\mu=m_1 m_2/M$ is the reduced mass, and $\nu \equiv \mu/M$ is the symmetric mass-ratio parameter. A symmetric binary ($m_1=m_2$) corresponds to $\nu=1/4$ and the test-particle limit is $\nu \to 0$. The total energy emitted during this regime can be estimated as the difference of the energy of the binary between infinity and the one at the last circular orbit (LCO). For a test-particle in the Schwarzschild background the LCO is located at $r_{\rm LCO}=6 G M/c^2$, its energy is $\sqrt{8/9} \mu c^2$, then:
\begin{equation}\label{eqn:energy-inspiral}
    \Delta E_{\rm insp} = \bigl(1-\sqrt{8/9}\bigr)\mu c^2.  
\end{equation}
%

\subsection{Merger regime}\label{sec:4b}

The GW spectrum of the merger regime is characterized by a GW burst \citep[see, e.g.,][]{1971PhRvL..27.1466D,2011LRR....14....6S,2015PhRvL.115i1101B}. Thus, to estimate whether this part of the signal contributes to the signal-to-noise ratio, it is sufficient to estimate the location of the GW burst in the frequency domain and its energy content. We recall that, in general, the merger regime is composed of plunge+merger+ringdown. The frequency range spanned by the GW burst is $\Delta f =f_{\rm qnm}-f_{\rm merger}$, where $f_{\rm merger}$ is the frequency at which the merger starts and $f_{\rm qnm}$ is the frequency of the ringing modes of the newly formed object after the merger, and the energy emitted is $\Delta E_{\rm merger}$. With these quantities defined, we can estimate the typical value of the merger regime spectrum as:
\begin{equation}
    \biggl(\frac{d E}{df_s}\biggr)_{\rm merger} \sim \frac{\Delta E_{\rm merger}}{\Delta f}.
    \label{eqn:spectrum-merger}
\end{equation}

Numerical relativity simulations \citep[e.g.][]{2011LRR....14....6S,2015PhRvL.115i1101B} show that finite size effects might end the inspiral regime before the LCO. After this point, the GW spectrum damps exponentially. For the case of NS-NS the merger starts in an orbit larger than the LCO, and for the case of a NS-BH, as we will see below, the merger can occur below the LCO making the spectrum similar to a BH-BH merger. When the merger occurs well before the LCO, there is no plunge. Therefore, the emitted energy will be less than the case when the plunge is present. We can therefore obtain an upper limit to $\Delta E_{\rm merger}$ by adopting the energy emitted during the plunge-merger-ringdown of a BH-BH merger \citep{1979ApJ...231..211D}
\begin{equation}\label{eq:deltaEPM}
\Delta E_{\rm merger} \approx 0.5\nu^2 M c^2.
\end{equation}
\begin{table*}
\caption{\label{tab:freqs}
Properties of the GW emission of S-GRFs, S-GRBs and U-GRBs. We have made the drastic {simplified} assumption that the binary evolution is only driven by GW emission, although it is manifest that the gravitational energy of the system in the merger phase is dominated by the radio, optical, X, gamma-ray and GeV emission (see Table~\ref{tab:rates}). This assumption is made with the only aim of establishing an absolute upper limit to the GW emission and its detectability under the most optimistic conditions. Column 1: GRB subclass. Column 2: energy emitted in GWs during the inspiral regime $\Delta E_{\rm insp}$ given by Eq.~(\ref{eqn:energy-inspiral}). Column 3: energy emitted in GWs during the merger regime (plunge+merger+ringdown) $\Delta E_{\rm merger}$ given by Eq.~(\ref{eq:deltaEPM}). Columns 4: GW frequency at merger. Column 5: GW frequency of the ringdown regime. Column 6: lowest cosmological redshift value $z_{\rm min}^{\rm obs}$ at which each subclass has been observed. Column 7: luminosity distance corresponding to $z_{\rm min}^{\rm obs}$, $d_{l_{\rm min}}$, estimated from Eq.~(\ref{eq:dl}). Columns 8--10: GW horizon calculated with the sensitivity of Advanced LIGO during the O1 and O2 runs and with the expected final sensitivity including LIGO-India (2022+), respectively. It can be seen that the current GW horizon is much smaller than the observed distances of GRBs, impeding a positive detection by Advanced LIGO. Only in the case of U-GRB (BH-SN) it is foreseen a possible detection during the run 2022+. See also Table~\ref{tab:summary}. We have used for S-GRFs (1.4+1.4)~$M_\odot$, for S-GRBs (2.0+2.0)~$M_\odot$ and, for U-GRBs (1.5+3.0)~$M_\odot$ and (1.5+10.0)~$M_\odot$ respectively for the \emph{out-states} of BdHNe and of BH-SN. Even if no U-GRB has yet been identified, we use here the values of $z_{\rm min}^{\rm obs}$ and $d_{l_{\rm min}}$ corresponding to the closest BdHN observed.}
\begin{ruledtabular}
\begin{tabular}{lccccccccc}
    & $\Delta E_{\rm insp}$ & $\Delta E_{\rm merger}$ & $f_{\rm merger}$ & $f_{\rm qnm}$ & $z_{\rm min}^{\rm obs}$   & $d_{l_{\rm min}}$  & \multicolumn{2}{c}{$d_{\rm GW}$ (Mpc)}\\
 & (erg) & (erg) & (kHz) & (kHz) &  & (Mpc)  &  O1 & O2 & 2022+ \\
\hline
S-GRF & $7.17\times 10^{52}$ & $1.60\times 10^{53}$ & 1.20 & 3.84 & $0.111$ &  508.70 & 90.51--181.02 & 181.02--271.52 & 452.54\\
S-GRB & $1.02\times 10^{53}$ & $2.28\times 10^{53}$ & 1.43 & 2.59 & $0.903$ & 5841.80 & 121.84--243.67 & 243.67--365.51 &  609.18\\
U-GRB & $1.02\times 10^{53}$ & $2.03\times 10^{52}$ & 0.98 & 2.30 & $0.169$ &  804.57 & 126.71--253.43 & 253.43--380.14 & 633.57\\
U-GRB (BH-SN) & $1.34\times 10^{53}$ & $1.35\times 10^{53}$ & 0.38 & 0.90 & $0.169$ &  804.57 & 197.86--395.71 & 395.71--593.57 & 989.28\\
\end{tabular}
\end{ruledtabular}
\end{table*}

To complete the estimate of the merger regime spectrum, we have to estimate the value of $\Delta f$ in the different cases of interest. 

\subsubsection{NS-NS merger}

The approach to the merger point, $r=r_{\rm merger}$, depends on the nature of the binary system. Typically, the merger is assumed to start at the point of maximum GW strain \citep[see, e.g.,][and references therein]{2015PhRvL.115i1101B}. However, since the transition from a binary system to a single merged object is not sharply definable, there can be found different definitions of the merger point in the literature \citep[see, e.g.,][]{2015PhRvD..92b4014K}. For our purpose it is sufficient to estimate the frequency at ``contact'', namely the frequency at a binary separation $r_{\rm contact}\approx r_1+r_2$ where $r_i$ is the radius of the $i$-component. This certainly sets a lower limit to the frequency at maximum strain at merger, i.e. $r_{\rm contact}\gtrsim r_{\rm merger}$. Thus, we adopt for these systems:
\begin{equation}
  f^{\rm NS-NS}_{\rm merger} \approx  f_{\rm contact}^{\rm NS-NS} = \frac{1}{\pi} \frac{c^3}{G M}\biggl[\frac{{\cal C}_1{\cal C}_2(1+q)}{{\cal C}_1+q{\cal C}_2}\biggr]^{3/2}, 
    \label{eqn:fNSNS}
\end{equation}
where $q=m_2/m_1$ is the mass-ratio which is related to the symmetric mass-ratio parameter by $\nu=q/(1+q)^2$, and ${\cal C}_i\equiv G m_i/c^2 r_i$ is the compactness of the $i$-component.

For a mass-symmetric NS-NS binary, we have that $f^{\rm NS-NS}_{\rm contact}\approx (1/\pi)(c^3/G){\cal C}_{\rm NS}^{3/2}/M$, where ${\cal C}_{\rm NS}\equiv {\cal C}_1={\cal C}_2$ is the compactness parameter of the initial NS. For example, for the NL3 EOS, the NS compactness lies in the range ${\cal C}_{\rm NS}\approx 0.14$--0.3 for a NS mass $1.4$--$2.8~M_\odot$ \citep[see, e.g.,][]{2015PhRvD..92b3007C}. Thus, using the same EOS we have, for an $M=(1.4+1.4)~M_\odot=2.8~M_\odot$ binary, $f^{\rm NS-NS}_{\rm contact}\approx$~1.34~kHz, and for an $M=(2.0+2.0)~M_\odot=4.0~M_\odot$ binary, $f^{\rm NS-NS}_{\rm contact}\approx$~1.43~kHz.

In the merger regime either a BH or a MNS can be formed. If the merger does not lead to a BH, the merger frequency is dominated by the frequency of the quasi-normal modes of the MNS formed. This frequency is of the order of 
\begin{equation}\label{eq:fqnmNS}
f^{\rm MNS}_{\rm qnm}\approx \frac{1}{\pi}\left(\frac{G M}{R^3}\right)^{1/2} = \frac{1}{\pi}\left(\frac{c^3}{G}\right)\frac{{\cal C}_{\rm MNS}^{3/2}}{M},
\end{equation}
where $R$ is the radius of the MNS and ${\cal C}_{\rm MNS}\equiv G M/(c^2 R)$ is its compactness. Thus, in the case of S-GRFs the value of $\Delta f$ is
\begin{eqnarray}\label{eq:deltafsgrf}
\Delta f_{\rm S-GRF} &\equiv& f^{\rm MNS}_{\rm qnm}-f^{\rm NS-NS}_{\rm contact}\nonumber \\ &\approx& ({\cal C}_{\rm MNS}^{3/2}-{\cal C}_{\rm NS}^{3/2}) \frac{c^3}{\pi G M}.
\end{eqnarray}

If the merger forms a BH, the merger frequency is dominated by the frequency of the quasi-normal modes of the BH formed, namely the GW-burst spectrum peaks at the frequency \citep{1971PhRvL..27.1466D,1972PhRvD...5.2932D}
\begin{equation}\label{eq:fqnmBH}
f^{\rm BH}_{\rm qnm}\approx \frac{0.32}{2\pi} \frac{c^3}{G M},
\end{equation}
i.e. $f_{\rm qnm}\approx 3.4$~kHz for a Schwarzschild BH of $3~M_\odot$. In the case of a rotating BH, namely a Kerr BH, the peak frequency shifts to higher values \citep{1980ApJ...239..292D}. Thus, the value of $f^{\rm BH}_{\rm qnm}$ given by Eq.~(\ref{eq:fqnmBH}) can be considered as a lower bound to the actual peak frequency. Thus, in the case of S-GRBs the value of $\Delta f$ is
\begin{eqnarray}\label{eq:deltafsgrb}
\Delta f_{\rm S-GRB}&\equiv& f^{\rm BH}_{\rm qnm}-f^{\rm NS-NS}_{\rm contact}\nonumber \\
&\approx& (0.16-{\cal C}_{\rm NS}^{3/2}) \frac{c^3}{\pi G M}.
\end{eqnarray}

In either case of BH or MNS formation, it is satisfied $f_{\rm qnm} > f_{\rm contact}$. It can be checked that the above frequency estimates are consistent with values obtained from full numerical relativity simulations \citep[see, e.g.,][]{1995PhRvD..52.2044A,2015PhRvL.115i1101B}.

\subsubsection{NS-BH merger}

For a NS-BH merger, the approach to merger is different since general relativistic effects avoid the objects to go all the way to the ``contact'' point following circular orbits. For example, let us assume $m_1=m_{\rm BH}\approx 3~M_\odot$ and $m_2=M_{\rm NS}\approx 1.5~M_\odot$, so $M= 1.5+3.0~M_\odot=4.5~M_\odot$. In this case $r_1=2 G m_{\rm BH}/c^2$ (for a Schwarzschild BH) and $r_2=G m_2/(c^2 {\cal C}_2)$, so $r_{\rm contact} \approx 3.3 G M/c^2$. Within the test-particle limit, the LCO around a Schwarzschild BH occurs at $r_{\rm LCO}=6 G m_{\rm BH}/c^2\approx 6 G M/c^2 > r_{\rm contact}$. Thus, we have that $r_{\rm contact}<r_{\rm LCO}$ which suggests that a NS-BH binary, similar to the case of a BH-BH one, can pass from the inspiral regime, to the plunge from $r_{\rm plunge}=r_{\rm LCO}$ to merger at $r_{\rm merger}\approx r_{\rm contact}$, to the ringing of the newly formed BH. At $r_{\rm plunge}$, the GW frequency is 
\begin{equation}\label{eq:fplunge}
f^{\rm NS-BH}_{\rm plunge}\approx \frac{1}{\pi}\left(\frac{G M}{r_{\rm LCO}^3}\right)^{1/2} = \frac{1}{\pi 6^{3/2}}\left(\frac{c^3}{G M}\right),
\end{equation}
and as in the previous case of BH formation from a NS-NS merger, the NS-BH post-merger GW spectrum will be dominated by frequencies given by Eq.~\eqref{eq:fqnmBH}. Namely, for the present example $f^{\rm NS-BH}_{\rm plunge} \approx 980$~Hz and $f^{\rm BH}_{\rm qnm}\approx 2.3$~kHz. 

{Thus, in the case of NS-BH merger (U-GRB subclass), the value of $\Delta f$ is}
\begin{equation}\label{eq:deltafugrb}
\Delta f_{\rm U-GRB} \equiv f^{\rm BH}_{\rm qnm}-f^{\rm NS-BH}_{\rm plunge} \approx 0.092 \frac{c^3}{\pi G M}.
\end{equation}

In the above analysis we have neglected the possibility that the NS can be tidally disrupted by the BH before it reaches $r=r_{\rm LCO}$.  The NS is disrupted by the BH if $r_{\rm LCO}<r_{\rm td}$, where $r_{\rm td}$ is the tidal disruption radius. The value of $r_{\rm LCO}$ and $r_{\rm td}$ for a NS-BH system depends both on the binary mass-ratio $q\equiv m_2/m_1 \leq 1$ and on the NS compactness ${\cal C}_{\rm NS}$ which depends, in turn, on the NS mass and EOS. Numerical simulations of NS-BH binary mergers adopting a polytropic EOS for the NS matter suggest $r_{\rm td}\approx 2.4 q^{-1/3} R_{\rm NS}$ and $r_{\rm LCO} \approx 6 G M/c^2 [1-0.44 q^{1/4}(1-3.54 {\cal C}_{\rm NS}]^{-2/3}$ \citep[see][and references therein]{2011LRR....14....6S}. The ratio $r_{\rm td}/r_{\rm LCO}$ is a decreasing function of the BH mass for given NS mass (but always close to unity). If we extrapolate these results to BH masses in the range (3--10)~$M_\odot$ and a NS of 1.5~$M_\odot$ obeying the NL3 EOS we have $r_{\rm LCO}<r_{\rm td}$ for $m_{\rm BH} \lesssim 6~M_\odot$ and $r_{\rm LCO}>r_{\rm td}$ otherwise. It is clear that the specific range of NS and BH masses for which there is tidal disruption is highly sensitive to the compactness of the NS, hence to the nuclear EOS, and thus more simulations using a wide set of updated nuclear EOS is needed to assess this issue. If tidal disruption occurs, the inspiral regime will cut-off at a GW frequency
\begin{equation}\label{eq:ftd}
f^{\rm NS-BH}_{\rm td}\approx \frac{1}{\pi}\left(\frac{G M}{r_{\rm td}^3}\right)^{1/2}.
\end{equation}
Since $r_{\rm td}$ is near $r_{\rm LCO}$ for our systems, and to not introduce further uncertainties in our estimates, we shall adopt that the inspiral regime of our NS-BH systems ends at the GW frequency given by Eq.~(\ref{eq:fplunge}).

\section{Characteristic strain and detectors sensitivity}\label{sec:5}

From Eqs.~(\ref{eqn:spectrum-inspiral}) and (\ref{eqn:spectrum-merger}) and with the knowledge of the energy released in GWs (\ref{eq:deltaEPM}) and the spanned frequencies in the merger regime (see Table~\ref{tab:freqs}), we can estimate the characteristic strain (\ref{eqn:h-char-gen}) which can be compared and contrasted with the strain noise of GW detectors.

\begin{figure*}
\centering
\includegraphics[width=0.8\hsize,clip]{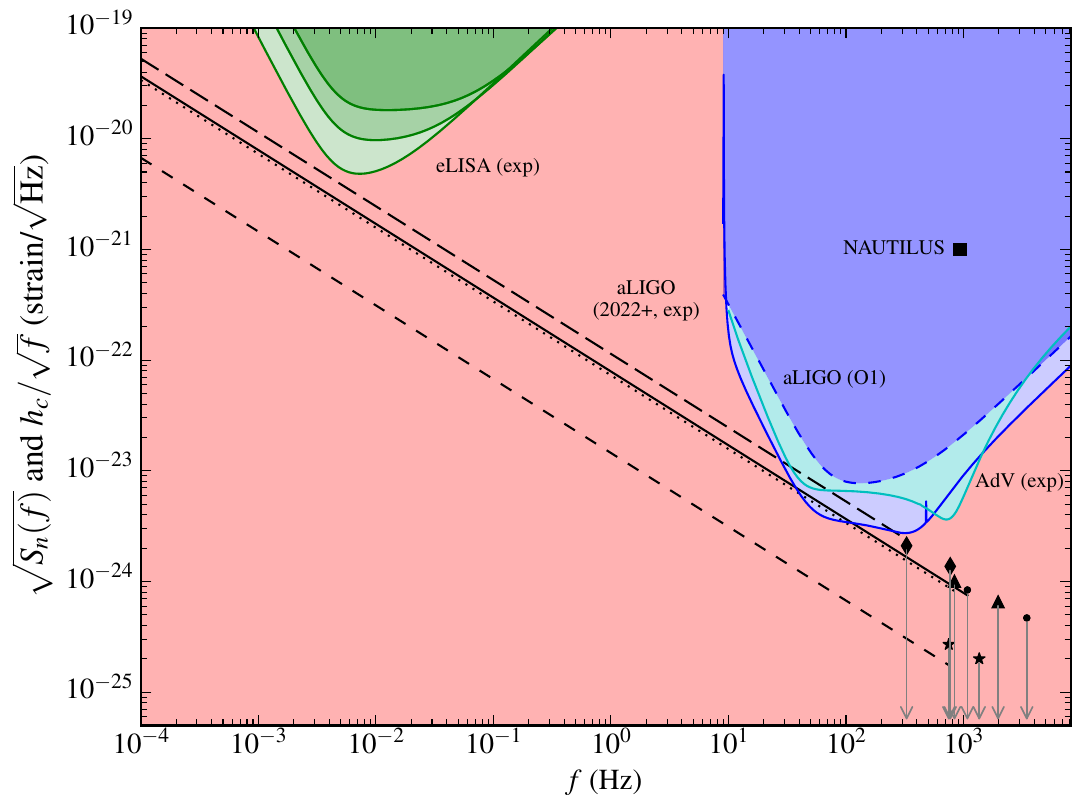}
\caption{Comparison of the signal's ASD $h_c/\sqrt{f}$ of S-GRFs, S-GRBs and U-GRBs with the noise's ASD $\sqrt{S_n(f)}$, where $S_n$ is the power spectrum density of the detector's noise of eLISA, of Advanced LIGO (aLIGO) and of the bar detector NAUTILUS. The red lines, from top to bottom, are the expected noise's ASD of the N2A1, N2A2 and N2A5 configurations of eLISA \citep{2016PhRvD..93b4003K}. The dashed and continuous red lines correspond to the noise's ASD respectively of Advanced LIGO O1 run (2015/2016) and of the expected Advanced LIGO 2022+ run \citep{2016LRR....19....1A}, and the cyan line is the expected noise's ASD of Advanced Virgo (AdV) BNS-optimized \citep{2016LRR....19....1A}. The filled square indicates the noise's ASD of the NAUTILUS resonant bar for a $1$~ms GW burst \citep{2006CQGra..23S..57A,2008CQGra..25k4048A}. The red filled area indicates the region of undetectability by any of the above instruments. We recall that in this plot the GW frequency is redshifted by a factor $1+z$ with respect to the source frame value, i.e.~$f = f_s/(1+z)$, for which we use the cosmological redshift and corresponding luminosity distance of the closest observed source of each subclass (see Table~\ref{tab:freqs}). The following three curves correspond to the inspiral regime of the coalescence: S-GRFs with $(1.4+1.4)~M_\odot$ (solid curve), S-GRBs with $(2.0+2.0)~M_{\odot}$ (short-dashed curve), U-GRB with $(1.5+3.0)~M_\odot$ (dotted curve) from \emph{out-states} of BdHNe, and U-GRB with $(1.5+10.0)~M_{\odot}$ (long dashed curve) from \emph{out-states} of BH-SNe. The dot, star, triangle and diamond correspond to $h_c$ in the merger regime for S-GRFs, S-GRBs, U-GRBs from \emph{out-states} of BdHNe, and U-GRBs from \emph{out-states} of BH-SNe, respectively. The first point is located at $f_{\rm merger}/(1+z)$ and the second at $f_{\rm qnm}/(1+z)$ (see Table~\ref{tab:freqs}). The down-arrows indicate that these estimates have to be considered as upper limits since we have assumed that all the energy release in the system goes in GWs, which clearly overestimates the GW energy output in view of the dominance of the electromagnetic emission (see Table~\ref{tab:summary}). We have also overestimated the GW energy in the merger regime by using Eq.~(\ref{eq:deltaEPM}) which is the expected GW energy emitted in the plunge+merger+ringdown phases of a BH-BH merger. For binary mergers involving NSs, as we have discussed in Sec.~\ref{sec:4}, the energy released in GWs must be necessarily lower than this value.}\label{fig:h_char}
\end{figure*}

Fig.~\ref{fig:h_char} shows the GW signal ASD produced by S-GRFs, S-GRBs and U-GRBs, obtained with the aid of Eq.~(\ref{eqn:h-char-gen}). In this figure we adopt: a $(1.4+1.4)~M_\odot$ $\nu$NS-NS merger for S-GRFs, a $(2.0+2.0)~M_\odot$ $\nu$NS-NS merger for S-GRBs, a $(1.5+3.0)~M_\odot$ $\nu$NS-BH merger for U-GRBs produced by \emph{out-states} of BdHNe, and a $(1.5+10.0)~M_\odot$ $\nu$NS-BH merger for U-GRBs produced by \emph{out-states} of BH-SNe. We have assumed in this plot that these sources are located at the closest luminosity distance $d_l$ at which each subclass has been observed (see Table~\ref{tab:freqs} for details). We show the noise ASD of Advanced LIGO in the current run (O1) and in the expected 2022+ run \citep{2016LRR....19....1A}; the expected noise ASD of Advanced Virgo \citep[BNS-optimized;][]{2016LRR....19....1A}; the expected noise ASD of the space-based interferometer eLISA for the N2A1, N2A2 and N2A5 configurations \citep[see, e.g.,][]{2016PhRvD..93b4003K}; and the noise ASD of the NAUTILUS bar detector for a $1$~ms GW burst \citep{2006CQGra..23S..57A,2008CQGra..25k4048A}. Narrow-band resonant bar detectors (such as ALLEGRO, AURIGA, EXPLORER, NAUTILUS and NIOBE) are sensitive within a bandwidth of $\sim 1$--10~Hz around the resonant frequency which is typically $f_0\sim 1$~kHz \citep[see, e.g., Table 2 in][for a summary of the properties of the bar detectors]{2004ARNPS..54..525C}. The bar detector with the wider bandwidth is NAUTILUS with a minimum strain spectral noise $\sqrt{S_n}=10^{-21}$~Hz$^{-1/2}$ at $f_0=935$~Hz and $\sqrt{S_n}\leq 10^{-20}$~Hz$^{-1/2}$ in a bandwidth $\sim 30$~Hz around $f_0$ \citep{2008CQGra..25k4048A}. This implies that a 1~ms GW burst would be detected by this instrument if it has a strain amplitude $h\gtrsim 3\times 10^{-19}$ \citep{2006CQGra..23S..57A,2008CQGra..25k4048A}.

From this figure we can conclude for the NS-NS and NS-BH binaries associated with S-GRFs, S-GRBs and U-GRBs:
\begin{enumerate}
\item
\emph{Before merging:} they transit, during their inspiral regime which spans the frequency range $f < f_{\rm merger}/(1+z)$ (see in Table~\ref{tab:freqs} the frequencies and redshift), first the eLISA frequency band to then enter the Advanced LIGO-Virgo ones in the final orbits prior the merging process (when $P_{\rm orb}<0.2$~s). The narrow bandwidth of the bar detectors does not cover these frequencies. For the adopted distances we see that the characteristic strain generated by all these sources is below the sensitivity of eLISA. S-GRFs are below the sensitivity of Advanced LIGO (O1), Advanced Virgo and NAUTILUS, but inside the sensitivity of Advanced LIGO (2022+). S-GRBs are below the sensitivity of Advanced LIGO (all runs), Advanced Virgo and NAUTILUS. U-GRBs from \emph{out-states} of BdHNe are below the sensitivity of Advanced LIGO (O1), Advanced Virgo and NAUTILUS, but inside the sensitivity of Advanced LIGO (2022+). U-GRBs from \emph{out-states} of BH-SNe are below the sensitivity of Advanced LIGO (O1) and NAUTILUS, inside the sensitivity of Advanced LIGO (2022+), and marginally inside the sensitivity of Advanced Virgo.
\item
\emph{Merging:} the merging regime, which expands frequencies from $f_{\rm contact}/(1+z)$ to $f_{\rm qnm}/(1+z)$ (see in Table~\ref{tab:freqs} the frequencies and redshift), is outside the eLISA frequency band but inside the Advanced LIGO-Virgo and bar detectors ones. The characteristic strain in this final merger phase $h\sim 10^{-24}$--$10^{-23}$ is, unfortunately, well below the sensitivity of all of them \citep[see, also,][for similar conclusions for Advanced LIGO]{2003apj...589..861k}.
\end{enumerate}

\begin{figure*}
\centering
\includegraphics[width=\hsize,clip]{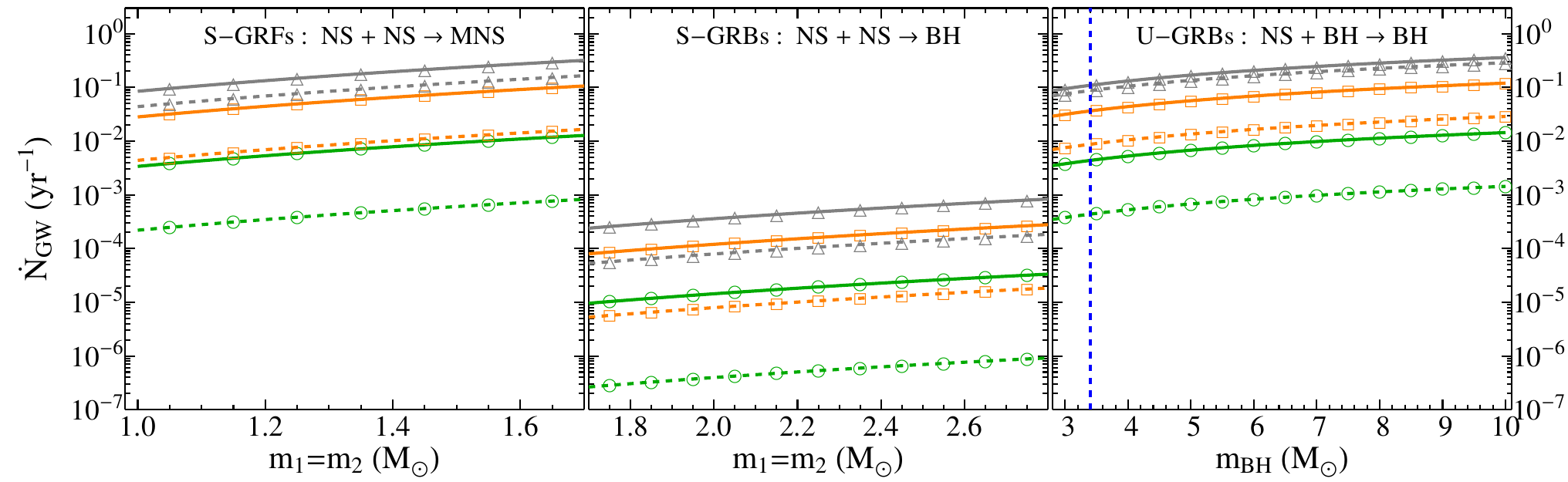}
\caption{Expected annual GW upper and lower bounds (the solid and the dashed lines, respectively) for the detections expected from S-GRFs (left panel), S-GRBs (middle panel), and U-GRBs (right panel), for three selected observational campaigns: 2015/2016 (O1: red curves with circles), 2017/2018 (O3: orange curve with squares), and $2022+$ (gray curve with triangles). The vertical red dashed line in the plot of U-GRBs separates $\nu$NS-BH binaries produced by BdHN (BH masses equal to the NS critical mass) and BH-SN (BH masses larger than the NS critical mass).}
\label{fig:GWdetGRBs}
\end{figure*}

From the above it can be seen that the most interesting instrument for the possible detection of the GW emission from binaries associated with GRBs is Advanced LIGO. Therefore, we estimate in the next section the expected detection rates by Advanced LIGO-Virgo (see Fig.~\ref{fig:GWdetGRBs} and Table~\ref{tab:summary}).

\begin{table*}
\caption{\label{tab:summary} Column 1: GRB subclass. Column 2: inferred number of GRBs per year in the entire Universe, $\dot{N}_{\rm GRB}$, for each GRB subclass \citep[see also Fig.~6 in][]{2016ApJ...832..136R}. Column 3: number of GRBs observed per year, $\dot{N}^{\rm obs}_{\rm GRB}$, obtained from the observations of $\gamma$-ray telescopes such as AGILE, BATSE, BeppoSAX, \emph{Fermi}, \emph{HETE-II}, \emph{INTEGRAL}, Konus/WIND and \emph{Swift}, in the indicated years of observations \citep[see also Tables~2--6 in][]{2016ApJ...832..136R}. Column 4: expected rate of GW detections by Advanced LIGO of all the GRB subclasses, computed for three selected observational campaigns: 2015/2016 (O1), 2016/2017 (O2), 2017/2018 (O3) and the one by the entire network at design sensitivity including LIGO-India (2022+). The binary component masses used here are the same of Table~\ref{tab:freqs}.
}
\begin{ruledtabular}
\begin{tabular}{llll}
GRB subclass & $\dot{N}_{\rm GRB}$ (yr$^{-1}$) & $\dot{N}^{\rm obs}_{\rm GRB}$ (yr$^{-1}$) & $\dot{N}_{\rm GW}$ (yr$^{-1}$) \\
\hline
XRFs &  $144$--$733$ & 1 (1997--2014) &  undetectable \\
\hline
BdHNe & $662$--$1120$ & 14 (1997--2014) & undetectable \\
\hline
BH-SN &  $\lesssim 662$--$1120$  & $\lesssim 14$ (1997--2014) & undetectable\\
\hline
S-GRFs & $58$--$248$ & 3 (2005--2014)& O1: 0.0001--0.002\\
& & & O2: 0.002--0.01\\
& & & O3: 0.008--0.05\\
& &  & 2022+: 0.1--0.2\\
\hline
S-GRBs  & $2$--$8$ & 1 (2006--2014)& O1: (0.1--3.1)$\times 10^{-6}$\\ 
& & & O2: (0.1--1.6)$\times 10^{-5}$\\
& & & O3: (0.6--7.8)$\times 10^{-5}$\\
& & & 2022+: (0.78--3.12)$\times 10^{-4}$\\
\hline
U-GRBs & $662$--$1120$ & -- & O1: (0.9--9)$\times 10^{-4}$\\
& & & O2: 0.001--0.005\\
& & & O3: 0.006--0.024\\
& & & 2022+: 0.076--0.094\\
\hline
U-GRBs (BH-SN) &  $\lesssim 662$--$1120$ & -- & O1: $\lesssim 0.00036$--0.0036\\
      &                          &  &  O2: $\lesssim 0.004$--0.018\\
      &                          &  &  O3: $\lesssim 0.02$--0.09\\
      &                          & & 2022+: $\lesssim 0.29$--0.36\\
\hline
GRFs  & 29-153 & 1 (2005--2014)& undetectable \\
\end{tabular}
\end{ruledtabular}
\end{table*}

\section{GW detection rate}\label{sec:6}

%
We assume a threshold for the Advanced LIGO-Virgo single detector $\rho_0 =8$ \citep{2016LRR....19....1A}. This minimum $\rho_0$ defines a maximum detection distance or GW horizon distance, which is denoted as $d_{\rm GW}$. This horizon corresponds to the most optimistic case when the binary is just above the detector and the binary plane is parallel to the detector plane, i.e. $\theta=\phi=\iota=0$ \citep{2012PhRvD..85l2006A}:
\begin{equation}
    d_{\rm GW} = \frac{2 A}{\rho_0} \biggl(\int_{0}^{\infty} \frac{f^{-7/3}}{S_n(f)}df\biggr)^{1/2},
    \label{eqn:d-horizon}
\end{equation}
where $A = 5/(24 \pi ^{4/3})^{1/2}(G M_c/c^3)^{5/6} c$. 
Since not all the sources are optimally aligned with the detector, the number of detected sources inside a sphere of radius $d_{\rm GW}$ will be a fraction $\mathcal{F}^3$ of the total. This fraction determines the so-called ``range'' of the detector, $\mathcal{R} = \mathcal{F}d_{\rm GW}$, where $\mathcal{F}^{-1} = 2.2627$ \citep[see][for details]{1993PhRvD..47.2198F}. In order to give an estimate of the annual number of detectable binaries associated with GRBs we use the \emph{search volume} as computed in \citep{2016LRR....19....1A}, $\mathcal{V}_s=V^{\rm GW}_{\rm max} \mathcal{T}$, where $V^{\rm GW}_{\rm max} = (4\pi/3) \mathcal{R}^3$ and  $\mathcal{T}$ is the observing time accounting for the detectors’ duty cycles. We use here the lower and upper values of $\mathcal{R}$ and $\mathcal{V}_s$ for a (1.4+1.4)~$M_\odot$ NS binary for the different observational campaigns reported in \citep{2016LRR....19....1A}: {2015/2016 (O1) with $\mathcal{R} = 40$--80~Mpc, $\mathcal{T} =3$~months, $\mathcal{V}_S = (0.5$--$4)\times10^5$~Mpc$^3$~yr; 2016/2017 (O2) with $\mathcal{R} = 80$--120~Mpc, $\mathcal{T} =6$~months, $\mathcal{V}_S = (0.6$--$2)\times10^6$~Mpc$^3$~yr;
2017/2018 (O3) with $\mathcal{R} = 120$--170~Mpc, $\mathcal{T}=9$~months, $\mathcal{V}_S = (3$--$10)\times10^6$~Mpc$^3$~yr, and the one by the entire network including LIGO-India at design sensitivity (2022+) with $\mathcal{R} = 200$~Mpc, $\mathcal{T}=1$~yr, $\mathcal{V}_S = 4\times10^7$~Mpc$^3$~yr.} We can use the above information for a (1.4+1.4)~$M_\odot$ binary and extrapolate it to other binaries with different masses using the property that $d_{\rm GW}$ scales with the chirp mass as $M^{5/6}_{c}$ (see Eq.~\ref{eqn:d-horizon}). {We show in Table~\ref{tab:freqs} GW horizon for a specific value of the binary component masses expected for S-GRFs, S-GRBs and U-GRBs (see section~\ref{sec:3b}).}

From the inferred occurrence rates $\rho_{\rm GRB}$ (not to be confused with signal-to-noise ratio $\rho$) summarized in Table~\ref{tab:rates}, we show in Fig.~\ref{fig:GWdetGRBs} the expected number of GW detections by Advanced LIGO-Virgo {for each observational campaign}
\begin{equation}\label{eq:Ngw}
\dot{N}_{\rm GW}=\rho_{\rm GRB} \mathcal{V}_s,
\end{equation}
for S-GRFs, S-GRBs, and U-GRBs {as a function of the binary component masses (see section~\ref{sec:3b})}. 

We compare and contrast in Table~\ref{tab:summary} for the GRB subclasses: the expected GW detection rate by Advanced LIGO-Virgo given by Eq.~(\ref{eq:Ngw}), $\dot{N}_{\rm GW}$, the inferred occurrence rate of GRBs, $\dot{N}_{\rm GRB}$, and the observed GRB rate from $\gamma$-ray telescopes (AGILE, BATSE, BeppoSAX, \emph{Fermi}, \emph{HETE-II}, \emph{INTEGRAL}, Konus/WIND and \emph{Swift}), simply estimated as $\dot{N}^{\rm obs}_{\rm GRB} = N^{\rm obs}_{\rm GRB}/T_{\rm obs}$ where $N^{\rm obs}_{\rm GRB}$ is the number of GRBs detected in the observing time $T_{\rm obs}$. The rate $\dot{N}_{\rm GRB}$ is obtained from the GRB specific rate through the reconstruction of the GRB luminosity function and the study of its evolution with the redshift \citep[see][{and Appendix~\ref{app:B}} for details]{2016ApJ...832..136R}. This estimate, therefore, is larger than $\dot{N}^{\rm obs}_{\rm GRB}$ which is limited to those events beyond the detector sensitivity threshold, falling inside its field of view and within its operational time.

\section{Conclusions}\label{sec:7}

Short and long GRBs have been divided into 7 subclasses according to their binary nature \citep{2016ApJ...832..136R}. We summarize in Table~\ref{tab:rates} their main physical properties characterizing the outcome of X-rays, gamma-rays, high-energy and ultra high-energy detectors, as well as their occurrence rate. Particularly important for the present work is the specification of the \emph{in-states} and \emph{out-states} of the GRB progenitors.

With the knowledge of the nature of the compact object binaries associated with each GRB subclass, {and the relevance of the NS structure and critical mass in Sec.~\ref{sec:2}, we introduce in Sec.~\ref{sec:3} the main ingredients for the computation of the GW emission and detectability for such systems.} We describe in Sec.~\ref{sec:4} the general properties of the GW emission during the inspiral and merger regimes of these binaries. We argue that S-GRFs, S-GRBs and U-GRBs are the GRB subclasses relevant for the GW analysis. It is manifest that the release of the gravitational energy of the system in the merger phase is dominated by the X-rays, gamma-rays and GeV emission (see Table~\ref{tab:rates}). In order to evaluate the GW emission we have made in this work the drastic {simplified} assumption that the binary evolution is only driven by GW emission. This assumption is of interest with the only aim of establishing an absolute upper limit and to check the detectability of the GW emission under this most optimistic condition. We compare and contrast in Sec.~\ref{sec:5} the GW characteristic strain amplitude produced by the inspiral and merger regimes with the strain noise of the broadband detectors eLISA, Advanced LIGO-Virgo as well as of the narrow-band resonant bar NAUTILUS. In order to do this we use the cosmological redshift and corresponding luminosity distance of the closest observed source of each subclass (see Table~\ref{tab:freqs}). We show that the inspiral regime is possibly detectable only by Advanced LIGO (see Table~\ref{tab:freqs} and Fig.~\ref{fig:h_char}) and the merger regime is undetectable by any of these instruments.

Therefore, in Sec.~\ref{sec:6} we assess quantitatively the GW detectability of the inspiral regime of S-GRFs, S-GRBs and U-GRBs only by Advanced LIGO. We recall that, following \citet{2016LRR....19....1A}, we adopt as the threshold for detectability a signal-to-noise ratio equal to 8. We present in Fig.~\ref{fig:GWdetGRBs} and Table~\ref{tab:summary} the expected detection rate of the GW emission. Four observational campaigns of Advanced LIGO are analyzed: O1 (2015/2016), O2 (2016/2017), O3 (2017/2018), and 2022+ namely the one by the entire network at design sensitivity including LIGO-India. In Table~\ref{tab:summary} we compare and contrast this rate with the occurrence rate of the GRB subclasses and their rate of observations by $\gamma$-ray telescopes.

Keeping the above in mind, we conclude for each GRB subclass:
\begin{enumerate}
\item[I.] {\bf XRFs:} their $\nu$NS-NS \emph{out-states} transit, during the inspiral regime which spans the frequency range $f < f_{\rm merger}/(1+z)$ (see Table~\ref{tab:freqs}), first the eLISA frequency band to then enter the Advanced LIGO-Virgo ones in the final orbits prior to the merging process (i.e. when $P_{\rm orb}<0.2$~s). Resonant bar detectors are not sensitive in this inspiral regime frequency range. The characteristic strain generated by these sources in the inspiral regime is below the sensitivity of eLISA. The merger regime, which expands frequencies from $f_{\rm contact}/(1+z)$ to $f_{\rm qnm}/(1+z)$ (see Table~\ref{tab:freqs}), is outside the eLISA frequency band but inside the frequency band of Advanced LIGO-Virgo and bar detectors. See Fig.~\ref{fig:h_char} for details. These $\nu$NS-NS mergers can lead either to S-GRFs or S-GRBs (see in IV and V below the conclusion about their GW detectability).

\item[II.] {\bf BdHNe:} their $\nu$NS-BH \emph{out-states} transit, during the inspiral regime which spans the frequency range $f < f_{\rm merger}/(1+z)$ (see Table~\ref{tab:freqs}), first the eLISA frequency band to then enter the Advanced LIGO-Virgo ones in the final orbits prior to the merging process (i.e. when $P_{\rm orb}<0.2$~s). Resonant bar detectors are not sensitive in this inspiral regime frequency range. The characteristic strain generated by these sources in the inspiral regime is below the sensitivity of eLISA. The merger regime, which expands frequencies from $f_{\rm contact}/(1+z)$ to $f_{\rm qnm}/(1+z)$ (see Table~\ref{tab:freqs}), is outside the eLISA frequency band but inside the frequency band of Advanced LIGO-Virgo and bar detectors. See Fig.~\ref{fig:h_char} for details. These $\nu$NS-BH mergers lead to U-GRBs (see in VI below the conclusion about their GW detectability).

\item[III.] {\bf BH-SN:} their $\nu$NS-BH \emph{out-states} transit, during the inspiral regime which spans the frequency range $f < f_{\rm merger}/(1+z)$ (see Table~\ref{tab:freqs}), first the eLISA frequency band to then enter the Advanced LIGO-Virgo ones in the final orbits prior to the merging process (i.e. when $P_{\rm orb}<0.2$~s). Resonant bar detectors are not sensitive in this inspiral regime frequency range. The characteristic strain generated by these sources in the inspiral regime is below the sensitivity of eLISA. The merger regime, which expands frequencies from $f_{\rm contact}/(1+z)$ to $f_{\rm qnm}/(1+z)$ (see Table~\ref{tab:freqs}), is outside the eLISA frequency band but inside the frequency band of Advanced LIGO-Virgo and bar detectors. See Fig.~\ref{fig:h_char} for details. These $\nu$NS-BH mergers lead to U-GRBs (see in VI below the conclusion about their GW detectability).

\item[IV.] {\bf S-GRFs:} the final orbits of the inspiral regime (when $P_{\rm orb}<0.2$~s) fall inside the frequency band of Advanced LIGO-Virgo and bar detectors. However, the GW energy output in the merger regime leads to a characteristic strain which is not sufficient to be detectable either by any of them. See Fig.~\ref{fig:h_char} for details. The inspiral regime is detectable for sources located at distances smaller than 181~Mpc for the O1 Advanced LIGO run and smaller than 452~Mpc for the 2022+ run (see Table~\ref{tab:freqs}). The closest S-GRF observed up to now is, however, located at 509~Mpc. See Table~\ref{tab:summary} for the expected GW detection rate.

\item[V.] {\bf S-GRBs:} the final orbits of the inspiral regime (when $P_{\rm orb}<0.2$~s) fall inside the frequency band of Advanced LIGO-Virgo and bar detectors. However, the GW energy output in the merger regime leads to a characteristic strain which is not sufficient to be detectable either by any of them. See Fig.~\ref{fig:h_char} for details. The inspiral regime is detectable for sources located at distances smaller than 244~Mpc for the O1 Advanced LIGO run and smaller than 609~Mpc for the 2022+ run (see Table~\ref{tab:freqs}). The closest S-GRB observed up to now is, however, located at 5842~Mpc. See Table~\ref{tab:summary} for the expected GW detection rate.

\item[VI.] {\bf U-GRBs:} the final orbits of the inspiral regime (when $P_{\rm orb}<0.2$~s) fall inside the frequency band of Advanced LIGO-Virgo and bar detectors. However, the GW energy output in the merger regime leads to a characteristic strain which is not sufficient to be detectable either by any of them. See Fig.~\ref{fig:h_char} for details. In the case of U-GRBs originating from the BdHN \emph{out-states}, the inspiral regime is detectable for sources located at distances smaller than 253~Mpc for the O1 Advanced LIGO run and smaller than 634~Mpc for the 2022+ run (see Table~\ref{tab:freqs}). In the case of U-GRBs originating from the BH-SN \emph{out-states}, the inspiral regime is detectable for sources at distances smaller than 396~Mpc for the O1 Advanced LIGO run and smaller than 989~Mpc for the 2022+ run (see Table~\ref{tab:freqs}). No U-GRB has yet been electromagnetically identified. The closest distance at which is located its possible progenitor, namely a BdHN, is 805~Mpc. See Table~\ref{tab:summary} for the expected GW detection rate.

\item[VII.] {\bf GRFs:} The tidal disruption of the WD by the NS produces a not detectable GW emission \citep[see, e.g.,][]{2009PhRvD..80b4006P}.

\end{enumerate}
 
We have recalled in the introduction that the simultaneous detection rate of GWs and GRBs have been estimated up to now in the literature only in the case of S-GRFs, in which no BH is formed but instead the merger of the two NSs leads to a MNS. Indeed, it can be seen that the recent GW detection rate estimated by \citet{2016JCAP...11..056P} of short bursts at Advanced LIGO design sensitivity (see Table 1 there), 0.04--15~yr$^{-1}$, is consistent with the one of S-GRFs estimated in this work, $\dot{N}_{\rm GW} = 0.1$--0.2~yr$^{-1}$ (see Table~\ref{tab:summary}). This represents the most favorable case for the possible GW detection by Advanced LIGO-Virgo of NS-NS merger which however does not lead to a BH formation but to a MNS.

We have given in this article, for the first time, a rate for the formation of BHs both in short and long bursts and this is of clear astrophysical relevance. Among such bursts producing a BH, the most favorable cases for GW detection are those from U-GRBs from BdHNe with $\dot{N}_{\rm GW} =0.08$--$0.09$~yr$^{-1}$ and those from BH-SN with $\dot{N}_{\rm GW} =0.3$--$0.4$~yr$^{-1}$ (see Table~\ref{tab:summary}). These NS-BH merging binaries were unknown in the literature and thus their occurrence and GW detection rates are a definite prediction of this work. 

Any detection by Advanced LIGO-Virgo of a NS-NS merger or a NS-BH merger will imply a drastic increase of the occurrence rate of events shown here and an examination of the consistency with GRB observations.

We have already given evidence on the unsuitability of the \emph{collapsar} model to explain the GRB observations in \citet{2018ApJ...852...53R}. We have published a classification on the ground of the current observations of 480 sources with known redshift \citep{2018ApJ...852...53R,2016ApJ...832..136R}, which is both necessary and sufficient, as of today, to cover all GRBs observed. As the number of sources will increase it is conceivable that the discovery of different systems will be observed and in that case we will be ready for their inclusion in additional subclasses within our classification scheme.

As we have mentioned the above are estimates based on most favorable conditions for GW emission and realistic $\dot{N}_{\rm GW}$ values will need the assessment of the GW to electromagnetic energy ratio which is necessarily smaller than unity from energy conservation.

{After the submission of this work, the LIGO-Virgo Collaboration announced the detection of the signal GW170817, and interpreted it as due to a NS-NS merger \citep{2017PhRvL.119p1101A}. As we have mentioned above, any possible GW detection of a NS-NS merger would imply a revision of its consistency with the inferences from GRB observations. It may then appear that our above conclusions of poor chance of detectability of NS-NS mergers by the Advanced LIGO-Virgo detector network are in tension with the detection of GW170817 during the O2 run. The association of GW170817 with GRB 170817A, a weak short-duration GRB observed by the Gamma-ray Burst Monitor (GBM) on board the Fermi-satellite \citep{2017ApJ...848L..13A,2017ApJ...848L..14G}, and followed-up in the optical bands \citep[e.g.][]{2017ApJ...848L..17C}, in the X-rays \citep[e.g.][]{2017ApJ...848L..25H} as well as by further gamma-rays facilities \citep[e.g.][]{2017ApJ...848L..15S}, allows us in the following to make an assessment on this issue.}

{First, we recall that GRB 170817A, with its isotropic energy emitted in gamma-rays of $E_{\rm iso}\approx 5\times 10^{46}$~erg \citep{2017ApJ...848L..14G} and peak luminosity of $(1.7\pm 0.1)\times 10^{47}$~erg~s$^{-1}$ \citep{2017arXiv171005851Z}, would belong to the S-GRF subclass if we assume it is produced in a NS-NS merger. On the other hand, we recall that our estimates of the local density rate of the GRB subclasses (see Table~\ref{tab:rates}), obtained from \citet{2016ApJ...832..136R}, are reliable for GRBs with luminosities higher than the lowest GRB luminosity in the subclass sample (see Appendix~\ref{app:B} for details). In the case of S-GRFs, we had identified GRB 050509B as the source with the lowest energetics: $E_{\rm iso}\approx 8.5\times 10^{48}$~erg (see Table 4 in \citealp{2016ApJ...832..136R}) and a peak luminosity $(1.1\pm 0.5)\times 10^{51}$~erg~s$^{-1}$ \citep{2005Natur.437..845F}. This implies that our predicted detention rates for the Advanced LIGO-Virgo detectors for S-GRFs are valid for sources with electromagnetic emission over the above values. Even a single observation of a close and underluminous source, as GRB 170817A,  would lead to an increase of the local density rate of this GRB subclass. Indeed, \citet{2017arXiv171005851Z} have recently estimated the increase in the local density rate when GRB 170817A is included in the sample of short bursts. Following a similar method to the one described in the Appendix~\ref{app:B}, they found that their previously estimated isotropic local density rate of (3.2--5.5)~Gpc$^{-3}$~yr$^{-1}$, obtained for sources with peak luminosities above $7\times 10^{49}$~erg~s$^{-1}$,\footnote{{This rate is consistent with the local density rate $\rho_{\rm S-GRFs}+\rho_{\rm S-GRBs}\approx \rho_{\rm S-GRFs} = (2.6$--5.0)~Gpc$^{-3}$~yr$^{-1}$ used in the present work; see Table~\ref{tab:rates} and \citet{2016ApJ...832..136R}.}} increases to a lower limit of (30--630)~Gpc$^{-3}$~yr$^{-1}$, for sources with peak luminosities above $1.7\times 10^{47}$~erg~s$^{-1}$, i.e. when GW170817 is included in the sample. The above range implies an increase of the local density rate by a factor $\sim 10$--100. It is then easy to check, using our Table~\ref{tab:summary}, that an increase of such a factor in the S-GRF density rate would imply a detection rate of (0.01--1)~yr$^{-1}$ for the O2 observational run, in agreement with the detection of GW170817.}

{In fact, the above isotropic density rate inferred by \citet{2017arXiv171005851Z} is consistent with the NS-NS observed merger rate of (320--4740)~Gpc$^{-3}$~yr$^{-1}$, inferred by the LIGO Collaboration with the detection of GW170817 \citep[see Sec.~V in][for details]{2017PhRvL.119p1101A}. This result implies that S-GRFs (or in general all short bursts) are not beamed or, if a beaming is assumed, the jet half-opening angle should be at least as large as 25--30$^\circ$.}

\section*{Acknowledgments}

We thank the referee for suggestions which improved the presentation of our results. M.K. acknowledges the support given by the International Relativistic Astrophysics Erasmus Mundus Joint Doctorate Program under the Grants 2013--1471, from EACEA of the European Commission. M.M. and J.A.R. acknowledge the partial support of the project No 3101/GF4 IPC-11 and the target program F.0679 of the Ministry of Education and Science of the Republic of Kazakhstan. C.C. and S.F. acknowledge INdAM-GNFM for support.


\appendix 
\section{IGC, Hypercritical Accretion, and Long GRBs}\label{app:A}

We give in this appendix details of the accretion process within the IGC scenario following \citet{2014ApJ...793L..36F,2015ApJ...812..100B,2015PhRvL.115w1102F,2016ApJ...833..107B}.

There are two main physical conditions for which hypercritical (i.e. highly super-Eddington) accretion onto the NS occurs in XRFs and BdHNe. The first is that the photons are trapped within the inflowing material and the second is that the shocked atmosphere on top the NS becomes sufficiently hot ($T\sim 10^{10}$~K) and dense ($\rho \gtrsim 10^6$~g~cm$^{-3}$) to produce a very efficient neutrino-antineutrino ($\nu\bar\nu$) cooling emission. In this way the neutrinos become the main responsible to release the energy gained by accretion, allowing hypercritical accretion to continue.

The first IGC simulations were performed in \citet{2014ApJ...793L..36F}, including: 1) realistic SN explosions of the CO$_{\rm core}$; 2) the hydrodynamics within the accretion region; 3) the simulated evolution of the SN ejecta up to their accretion onto the NS. \citet{2015ApJ...812..100B} then estimated the amount of angular momentum carried by the SN ejecta and how much is transferred to the NS companion by accretion. They showed that the SN ejecta can circularize for a short time and form a disc-like structure surrounding the NS before being accreted. The evolution of the NS central density and rotation angular velocity (the NS is spun up by accretion) was computed from full numerical solutions of the axisymmetric Einstein equations. The unstable limits of the NS are set by the mass-shedding (or Keplerian) limit and the critical point of gravitational collapse given by the secular axisymmetric instability \citep[see, e.g.,][for details]{2015ApJ...812..100B}.

\begin{figure}
\centering
\includegraphics[width=0.4\hsize,clip]{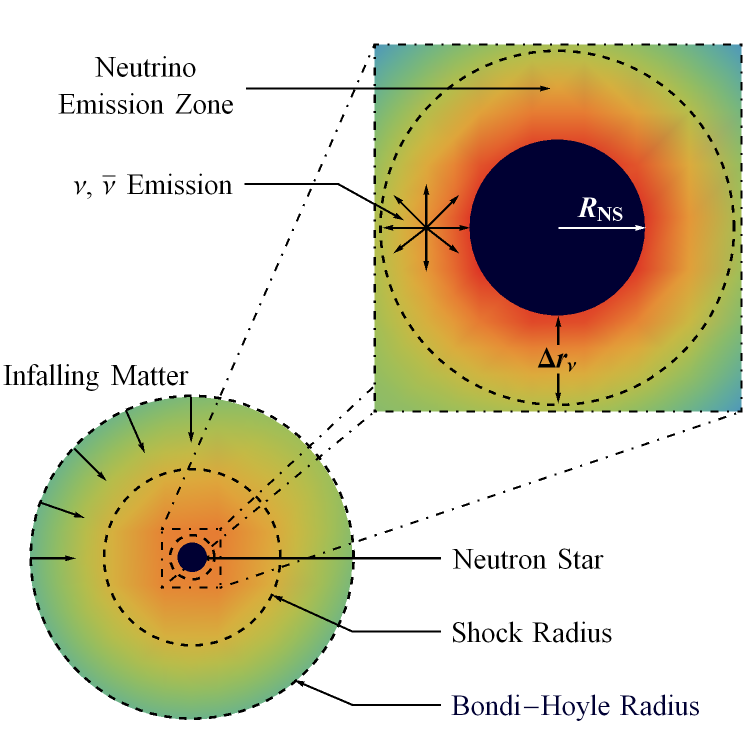}
\caption{Scheme of the IGC scenario: the CO$_{\rm core}$ undergoes SN explosion, the NS accretes part of the SN ejecta and then reaches the critical mass for gravitational collapse to a BH, with consequent emission of a GRB. The SN ejecta reach the NS Bondi-Hoyle radius and fall toward the NS surface. The material shocks and decelerates while it piles over the NS surface. At the neutrino emission zone, neutrinos take away most of the gravitational energy gained by the matter infall. The neutrinos are emitted above the NS surface in a region of thickness $\Delta r_\nu$ about half the NS radius that allow the material to reduce its entropy to be finally incorporated to the NS. For further details and numerical simulations of the above process see \citet{2014ApJ...793L..36F,2015ApJ...812..100B,2016ApJ...833..107B}.
}\label{fig:AccretionEsqueme}
\end{figure}

The accretion rate of the SN ejecta onto the NS is given by:
\begin{equation}\label{eq:BondiMassRate_definition}
\begin{array}{rclcrcl}
\dot{M}_B(t)&=&\pi\rho_{\rm ej}R_{{\rm cap}}^2\sqrt{v_{{\rm rel}}^2+c_{\rm s,ej}^2}, \qquad R_{{\rm cap}}(t)&=&\cfrac{2 G M_{{\rm NS}}(t)}{v_{{\rm rel}}^2+c_{{\rm s,ej}}^2},
 \end{array}
\end{equation}
where $G$ is the gravitational constant, $\rho_{\rm ej}$ and $c_{{\rm s,ej}}$ are the density and sound speed of the ejecta, $R_{\rm cap}$ and $M_{\rm NS}$ are the NS gravitational capture radius (Bondi-Hoyle radius) and gravitational mass, and $v_{{\rm rel}}$ the ejecta velocity relative to the NS: $\vec{v}_{{\rm rel}}=\vec{v}_{{\rm orb}}-\vec{v}_{{\rm ej}}$; $|\vec{v}_{\rm orb}|=\sqrt{G(M_{\rm core}+M_{\rm NS})/a}$, and $\vec{v}_{\rm ej}$ is the velocity of the supernova ejecta (see Fig.~\ref{fig:AccretionEsqueme}).

Numerical simulations of the SN explosions suggest the adopted homologous expansion of the SN, i.e.~$v_{\rm ej}(r,t)=n r/t$, where $r$ is the position of each layer from the SN center and $n$ is the expansion parameter. The density evolves as 
\begin{equation}
\rho_{\rm ej}(r,t)=\rho_{\rm ej}^0(r/R_{\rm star}(t),t_0)\frac{M_{\rm env}(t)}{M_{\rm env}(0)}\left(\frac{R_{\rm star}(0)}{R_{\rm star}(t)}\right)^3,
\end{equation}
where $M_{\rm env}(t)$ the mass of the CO$_{\rm core}$ envelope, $R_{\rm star}(t)$ is the radius of the outermost layer, and $\rho_{\rm ej}^0$ is the pre-SN CO$_{\rm core}$ density profile; $\rho_{\rm ej}(r,t_0)= \rho_{\rm core} (R_{\rm core}/r)^m$, where $\rho_{\rm core}$, $R_{\rm core}$ and $m$ are the profile parameters obtained from numerical simulations. Typical parameters of the CO$_{\rm core}$ mass are (3.5--9.5)~$M_\odot$ corresponding to $(15$--$30)~M_\odot$ zero-age-main-sequence (ZAMS) progenitors \citep[see][for details]{2014ApJ...793L..36F,2015ApJ...812..100B}. The binary period is limited from below by the request of having no Roche lobe overflow by the CO$_{\rm core}$ before the SN explosion \citep{2014ApJ...793L..36F}. For instance, for a CO$_{\rm core}$ of 9.5~$M_\odot$ forming a binary system with a 2~$M_\odot$ NS, the minimum orbital period allowed by this condition is $P_{\rm  min}\approx 5$~min. For these typical binary and pre-SN parameters, Eq.~(\ref{eq:BondiMassRate_definition}) gives accretion rates $10^{-4}$--$10^{-2} M_\odot$~s$^{-1}$.

We adopt an initially non-rotating NS companion so its exterior spacetime at time $t=0$ is described by the Schwarzschild metric. The SN ejecta approach the NS with specific angular momentum, $l_{\rm acc}=\dot{L}_{\rm cap}/\dot{M}_B$, circularizing at a radius $r_{\rm circ}\geq r_{\rm lco}$  if $l_{\rm acc} \geq l_{\rm lso}$ with $r_{\rm lco}$ the radius of the last circular orbit (LCO). For a non-rotating NS $r_{\rm lco}=6 G M_{\rm NS}/c^2$ and $l_{\rm lco}=2\sqrt{3} G M_{\rm NS}/c$. For typical parameters, $r_{\rm circ}/r_{\rm lco}\sim 10$--$10^3$. 

The accretion onto the NS proceeds from the radius $r_{\rm in}$. The Ns mass and angular angular momentum evolve as \citep{2015ApJ...812..100B,2017PhRvD..96b4046C}:
\begin{equation}\label{eq:AngMom}
\dot{M}_{\rm NS}=\left(\frac{\partial M_{\rm NS}}{\partial M_b}\right)_{J_{\rm NS}}\dot{M}_b+\left(\frac{\partial M_{\rm NS}}{\partial J_{\rm NS}}\right)_{M_b}\dot{J}_{\rm NS},\qquad \dot{J}_{\rm NS}=\xi \, l(r_{\rm in})\dot{M}_{\rm B},
\end{equation}
where $M_b$ is the NS baryonic mass, $l(r_{\rm in})$ is the specific angular momentum of the accreted material at $r_{\rm in}$, which corresponds to the angular momentum of the LCO, and $\xi\leq 1$ is a parameter that measures the efficiency of angular momentum transfer. In this picture we have $\dot{M}_b = \dot{M}_B$.

For the integration of Eqs.~(\ref{eq:BondiMassRate_definition}) and (\ref{eq:AngMom}) we have to supply the values of the two partial derivatives in Eq.~(\ref{eq:AngMom}). They are obtained from the relation of the NS gravitational mass, $M_{\rm NS}$, with $M_b$ and $J_{\rm NS}$, namely from the knowledge of the NS binding energy. For this we use the general relativistic calculations of rotating NSs presented in \citet{2015PhRvD..92b3007C}. They show that, independent on the nuclear EOS, the following analytical formula represents the numerical results with sufficient accuracy (error $<2\%$):
\begin{equation}\label{eq:MbMnsjns}
\frac{M_b}{M_\odot}=\frac{M_{\rm NS}}{M_\odot}+\frac{13}{200}\left(\frac{M_{\rm NS}}{M_\odot}\right)^2\left(1-\frac{1}{137}j_{\rm NS}^{1.7}\right),
\end{equation}
where $j_{\rm NS}\equiv cJ_{\rm NS}/(G M_\odot^2)$. 

In the accretion process the NS gains angular momentum and therefore spin up. To evaluate the amount of angular momentum transferred to the NS at any time we include the dependence of the LCO specific angular momentum as a function of $M_{\rm NS}$ and $J_{\rm NS}$. For corotating orbits the following relation is valid for the NL3, TM1 and GM1 EOS \citep{2017PhRvD..96b4046C,2015ApJ...812..100B}:
\begin{equation}
l_{\rm lco}= \frac{G M_{\rm NS}}{c}\left[2 \sqrt{3} - 0.37 \left(\frac{j_{\rm NS}}{M_{\rm NS}/M_\odot}\right)^{0.85}\right].
\end{equation}

The NS continues to accrete until an instability limit is reached or up to when all the SN ejecta overcomes the NS Bondi-Hoyle region. We take into account the two main instability limits for rotating NSs: the mass-shedding or Keplerian limit and the secular axisymmetric instability limit. The latter defines critical NS mass. For the aforementioned nuclear EOS, the critical mass is approximately given by \citep{2015PhRvD..92b3007C}:
\begin{equation}\label{eq:Mcrit}
M_{\rm NS}^{\rm crit}=M_{\rm NS}^{J=0}(1 + k j_{\rm NS}^p),
\end{equation}
where $k$ and $p$ are EOS-dependent parameters (see Table~\ref{tb:StaticRotatingNS}). These formulas fit the numerical results with a maximum error of 0.45\%.
\begin{table}
\centering
\caption{Critical NS mass in the non-rotating case and constants  $k$ and $p$ needed to compute the NS critical mass in the non-rotating case given by Eq.~(\ref{eq:Mcrit}). The values are given for the NL3, GM1 and TM1 EOS.}
{\begin{tabular}{@{}cccc@{}} 
\toprule
EOS  &  $M_{\rm crit}^{J=0}$~$(M_{\odot})$ & $p$&$k$ \\\colrule
NL3 & $2.81$ & $1.68$ & $0.006$\\
GM1 & $2.39$ & $1.69$ & $0.011$\\
TM1 & $2.20$ & $1.61$ & $0.017$\\
\botrule
\end{tabular}
}
\label{tb:StaticRotatingNS}
\end{table}

\subsection{Most recent simulations of the IGC process}\label{app:A1}

Additional details and improvements of the hypercritical accretion process leading to XRFs and BdHNe were presented in \citet{2016ApJ...833..107B}. Specifically:
\begin{enumerate}
\item
The density profile included finite size/thickness effects and additional CO$_{\rm core}$ progenitors leading to different SN ejecta masses were considered.
\item
In \citet{2015ApJ...812..100B} the maximum orbital period, $P_{\rm max}$, over which the accretion onto NS companion is not sufficient to bring it to the critical mass, was inferred. Thus, binaries with $P > P_{\rm max}$ lead to XRFs while the ones with $P\lesssim P_{\rm max}$ lead to BdHNe. \citet{2016ApJ...833..107B} extended the determination of $P_{\rm max}$ for all the possible initial values of the NS mass. They also examined the outcomes for different values of the angular momentum transfer efficiency parameter.
\item
It was estimated the expected luminosity during the process of hypercritical accretion for a wide range of binary periods covering both XRFs and BdHNe.
\item
It was shown that the presence of the NS companion originates asymmetries in the SN ejecta (see, e.g., Fig.~6 in \citealp{2016ApJ...833..107B}). The signatures of such asymmetries in the X-ray emission was there shown in the specific example of XRF 060218.
\end{enumerate}

\subsection{Hydrodynamics and neutrino emission in the accretion region}\label{app:A2}

The accretion rate onto the NS can be as high as $\sim 10^{-2}$--$10^{-1}~M_{\odot}$~s$^{-1}$. For such accretion rates:
\begin{enumerate}
\item 
The magnetic pressure is much smaller than the random pressure of the infalling material, therefore the magnetic-field effects on the accretion process are negligible \citep{1996ApJ...460..801F,2012ApJ...758L...7R}.
\item 
The photons are trapped within the infalling matter, hence the Eddington limit does not apply and hypercritical accretion occurs. The trapping radius is defined by \citep{1989ApJ...346..847C}: $r_{\rm trapping}={\rm min}\{\dot{M}_B\kappa/(4\pi c),R_{\rm cap}\}$, where $\kappa$ is the opacity. \citet{2014ApJ...793L..36F} estimated a Rosseland mean opacity of $\approx 5\times 10^3$~cm$^2$~g$^{-1}$ for the CO$_{\rm cores}$. This, together with our typical accretion rates, lead to $\dot{M}_B\kappa/(4\pi c)\sim 10^{13}$--$10^{19}$~cm. This radius is much bigger than the Bondi-Hoyle radius. 
\item 
The above condition, and the temperature-density values reached on top the NS surface, leads to an efficient neutrino cooling which radiates away the gain of gravitational energy of the infalling material \citep{1972SvA....16..209Z,1973PhRvL..31.1362R,1996ApJ...460..801F,2012ApJ...758L...7R,2014ApJ...793L..36F}.
\end{enumerate}

\subsubsection{Convective instabilities}

The accretion shock moves outward as the material piles onto the NS. Since the post-shock entropy is inversely proportional to the shock radius position the NS atmosphere is unstable with respect to Rayleigh-Taylor convection at the beginning of the accretion process. Such instabilities might drive high-velocity outflows from the accreting NS \citep{2006ApJ...646L.131F,2009ApJ...699..409F}. The entropy at the base of the atmosphere is \citep{1996ApJ...460..801F}: 
\begin{equation}
S_{\rm bubble} \approx 16\left(\frac{1.4\,M_\odot}{M_{\rm NS}}\right)^{-7/8}\left(\frac{M_\odot\,{\rm s}^{-1}}{\dot{M}_{\rm B}}\right)^{1/4}\left(\frac{10^6\, {\rm cm}}{r}\right)^{3/8}\,k_B/{\rm nucleon},
\end{equation}
The material expands and cools down adiabatically, i.e.~$T^3/\rho$ = constant. In the case of a spherically symmetric expansion, $\rho \propto 1/r^3$ and $k_B T_{\rm bubble}=195\, S_{\rm bubble}^{-1}\left(10^6\, {\rm cm}/r\right)$~MeV. In the more likely case that the material expand laterally we have \citep{2009ApJ...699..409F}: $\rho \propto 1/r^2$, i.e. $T_{\rm bubble} = T_0 (S_{\rm bubble}) \left(r_0/r\right)^{2/3}$, where $T_0(S_{\rm bubble})$ is obtained from the above equation at $r=r_0\approx R_{\rm NS}$. This implies a bolometric blackbody flux at the source from the rising bubbles:
\begin{equation}\label{eq:Lbubble}
F_{\rm bubble} \approx 2\times 10^{40} \left(\frac{M_{\rm NS}}{1.4\,M_\odot} \right)^{-7/2}\left( \frac{\dot{M}_{\rm B}}{M_\odot\,{\rm s}^{-1}} \right)\left( \frac{R_{\rm NS}}{10^6\,{\rm cm}} \right)^{3/2}\left(\frac{r_0}{r}\right)^{8/3}\,{\rm erg\,s}^{-1}{\rm cm}^{-2},
\end{equation}
where $\sigma$ is the Stefan-Boltzmann constant.

The above thermal emission has been shown  \citep{2014ApJ...793L..36F} to be a plausible explanation of the early ($t\lesssim 50$~s) X-ray emission observed in some GRBs. In the specific example of GRB 090618 \citep{2012A&A...543A..10I,2012A&A...548L...5I}, adopting an accretion rate of $10^{-2}~M_\odot$~s$^{-1}$, the bubble temperature drops from 50~keV to 15~keV while expanding from $r\approx 10^9$~cm to $6\times 10^9$~cm.

\subsubsection{Neutrino emission and effective accretion rate}

Temperatures $k_B T\sim 1$--10~MeV and densities $\rho\gtrsim 10^{6}$~g~cm$^{-3}$ develop near the NS surface during the accretion process. Under these conditions, $e^+e^-$annihilation into $\nu\bar\nu$ pairs becomes the dominant neutrino emission process in the accretion region \citep[see][for details]{2016ApJ...833..107B}. The effective accretion rate onto the NS can be estimated as \citep[e.g.]{1996ApJ...460..801F}: $\dot{M}_{\rm eff} \approx \Delta M_\nu (L_\nu/E_\nu)$, where $\Delta M_\nu$ and $L_\nu$ are the mass and neutrino luminosity in the emission region; $E_\nu$ is half the gravitational potential energy gained by the material falling from infinity to a distance $\Delta r_\nu$ from the NS surface. $\Delta r_\nu$ is the thickness of the neutrino emitting region which is approximately given by the temperature scale height ($\Delta r_\nu \approx 0.6 R_{\rm NS}$). Since $L_\nu\approx 2\pi R_{\rm NS}^2\Delta r_\nu \epsilon_{\rm e^{-}e^{+}}$ with $\epsilon_{\rm e^{-}e^{+}}$ the $e^+e^-$ pair annihilation process emissivity, and $E_\nu=(1/2) G M_{\rm NS} \Delta M_\nu/(R_{\rm NS}+\Delta r_\nu)$, for $M_{\rm NS}=1.4~M_\odot$ one obtains $\dot{M}_{\rm eff} \approx 10^{-9}$--$10^{-1}~M_\odot$~s$^{-1}$ for $k_B T = 1$--10~MeV.
  
\subsection{Accretion luminosity}\label{app:A3}

The energy release in a time-interval $dt$, when an amount of mass $dM_b$ with  angular momentum $l \dot{M}_b$ is accreted, is:
\begin{equation}
L_{\rm acc}= (\dot{M}_b - \dot{M}_{\rm NS})c^2 =\dot{M}_b c^2 \left[1-\left(\frac{\partial M_{\rm NS}}{\partial J_{\rm NS}}\right)_{M_b}\,l -\left(\frac{\partial M_{\rm NS}}{\partial M_b}\right)_{J_{\rm NS}}\right].
\label{eq:DiskLuminosity}
\end{equation}
This is the amount of gravitational energy gained by the matter by infalling to the NS surface that is not spent in NS gravitational binding energy. The total energy release in the time interval from $t$ to $t+dt$, $\Delta E_{\rm acc} \equiv \int L_{\rm acc}dt$, is given by the NS binding energy difference between its initial and final state. The typical luminosity is $L_{\rm acc}\approx \Delta E_{\rm acc}/\Delta t_{\rm acc}$, where $\Delta t_{\rm acc}$ is the duration of the accretion process.

The value of $\Delta t_{\rm acc}$ is approximately given by the flow time of the slowest layers of the SN ejecta to the NS companion position. If we denote the velocity of these layers by $v_{\rm inner}$, we have $\Delta t_{\rm acc}\sim a/v_{\rm inner}$, where $a$ is the binary separation. For $a\sim 10^{11}$~cm and $v_{\rm inner}\sim 10^8$~cm~s$^{-1}$, $\Delta t_{\rm acc}\sim 10^3$~s. For shorter separations, e.g.~$a\sim 10^{10}$~cm ($P\sim 5$~min), $\Delta t_{\rm acc}\sim 10^2$~s. For a binary with $P=5$~min, the NS accretes $\approx 1~M_\odot$ in $\Delta t_{\rm acc}\approx 100$~s. From Eq.~(\ref{eq:MbMnsjns}) one obtains that the binding energy difference of a $2~M_\odot$ and a $3~M_\odot$ NS, is $\Delta E_{\rm acc}\approx 13/200 (3^2-2^2)~M_\odot c^2\approx 0.32~M_\odot c^2$. This leads to $L_{\rm acc}\approx 3\times 10^-{3}~M_\odot c^2 \approx 0.1 \dot{M_b} c^2$. The accretion power can be as high as $L_{\rm acc}\sim 0.1 \dot{M_b} c^2\sim 10^{47}$--$10^{51}$~erg~s$^{-1}$ for accretion rates in the range $\dot{M_b}\sim 10^{-6}$--$10^{-2}~M_\odot$~s$^{-1}$.

\subsection{Possible evolutionary scenario for CO$_{\rm core}$-NS binary formation}\label{app:A4}

Two independent communities have introduced a new evolutionary scenario for the formation of compact-object binaries (NS-NS or NS-BH). After the collapse of the primary star forming a NS, the binary undergoes mass-transfer episodes finally leading to the ejection of both the hydrogen and helium shells of the secondary star. These processes leads naturally to a binary composed of a CO$_{\rm core}$ and a NS companion. In the X-ray binary and SN communities these systems are called ``ultra-stripped'' binaries \citep[see, e.g.,]{2015MNRAS.451.2123T}. These
systems are expected to comprise 0.1-–1\% of the total SNe \citep{2013ApJ...778L..23T}. 

In the above studies most of the binaries have orbital periods in the range $3\times 10^3$--$3\times 10^5$~s which are longer with respect to the short periods expected in the BdHN scenario. The formation of the CO$_{\rm core}$-NS binaries leading to BdHNe might be a subset of the ultra-stripped binaries. In such subset the conditions of the initial orbital separation and CO$_{\rm core}$ mass must be such to lead to final orbital periods in the range 100-–1000~s. Assuming a SN rate of $2\times 10^4$~Gpc$^{-3}$~yr$^{-1}$ \citep{2007ApJ...657L..73G}, the ultra-stripped binaries would have a rate of 20--200~Gpc$^{-3}$~yr$^{-1}$, and thus BdHNe, with a rate of $\sim 1$~Gpc$^{-3}$~yr$^{-1}$ (see Table~\ref{tab:rates} and \citealp{2016ApJ...832..136R}), might be produced by the 0.5--5\% of the ultra-stripped binary population.

\subsection{Post-Explosion Orbits and NS-BH Binaries formation}\label{app:A5}

The SN explosion leaves as a central remnant the $\nu$NS, while the NS companion might lead, for sufficient accretion rates, to the formation of a BH. We examined in \citet{2015PhRvL.115w1102F} the question if BdHNe can indeed form NS-BH binaries or, on the contrary, they are disrupted by the SN explosion.

Most of the typical binaries become unbound during a SN explosion owing to the amount of mass loss and momentum imparted (kick) to the $\nu$NS in the explosion. Assuming an instantaneous explosion, the binary is disrupted if half of the binary mass is lost. For this reason the fraction of massive binaries that can produce double compact-object binaries might be as low as $\sim$0.001--1\% \citep{1999ApJ...526..152F,2012ApJ...759...52D,2014LRR....17....3P}. Indeed, this is consistent with our estimated GRB local observed rates: we have shown in section~\ref{sec:3a} that the NS-NS population leading to short-bursts can be explained as being descendant from the CO$_{\rm core}$-NS if $\sim 1\%$ of them remain bound after the SN explosion.

Assuming instantaneous mass loss, the post-explosion semi-major axis is \citep{1983ApJ...267..322H}: 
\begin{equation}
\frac{a}{a_0}=\frac{M_0 - \Delta M}{M_0 - 2 a_0 \Delta M/r},
\end{equation}
where $a_0$ and $a$ are the initial and final semi-major axes respectively, $M_0$ is the (initial) binary mass, $\Delta M$ is the change of mass (in this case the amount of mass loss), and $r$ is the orbital separation before the explosion. For circular orbits, the system is unbound if it loses half of its mass. For the very tight BdHNe, however, additional effects have to be taken into account to determine the fate of the binary.

The shock front in a SN moves at roughly $10^4$~km~s$^{-1}$, but the denser, lower-velocity ejecta, can move at velocities as low as $10^2$--$10^3$~km~s$^{-1}$ \citep{2014ApJ...793L..36F}. This implies that the SN ejecta overcomes a NS companion in a time 10--1000~s. For wide binaries this time is a small fraction of the orbital period and the ``instantaneous'' mass-loss assumption is perfectly valid. BdHNe have instead orbital periods as short as 100--1000~s, hence the instantaneous mass-loss approximation breaks down.

We recall the specific examples studied in \citet{2015PhRvL.115w1102F}: close binaries in an initial circular orbit of radius $7\times10^9$~cm, CO$_{\rm core}$ radii of $(1$--$4) \times 10^9$~cm with a 2.0~$M_\odot$ NS companion. The CO$_{\rm core}$ leaves a central 1.5~$M_\odot$ NS, ejecting the rest of the core. The NS leads to a BH with a mass equal to the NS critical mass. For these parameters it was there obtained that even if 70\% of the mass is lost the binary remains bound providing the explosion time is of the order of the orbital period ($P=180$~s) with semi-major axes of less than $10^{11}$~cm. 

The tight $\nu$NS-BH binaries produced by BdHNe will, in due time, merge owing to the emission of GWs. For the above typical parameters the merger time is of the order of $10^4$~yr, or even less. We expect little baryonic contamination around such merger site since this region has been cleaned-up by the BdHN. These conditions lead to a new family of sources which we have called ultrashort GRBs, U-GRBs.

\section{Local density rate of GRB subclasses}\label{app:B}

We recall now the method used in \citet{2016ApJ...832..136R} to estimate, for each GRB subclass, the local observed density rates that we use in this work. This is defined by the convolution of the luminosity function, that tells us the fraction of bursts with isotropic equivalent luminosities in the interval $\log L$ and $\log L+d \log L$, and the cosmic GRB occurrence rate, that tells us the number of sources at different redshifts. These functions depend on \textit{a priori} assumptions and some investigations have been carried out in the literature: for long bursts \citep[e.g.][]{Soderberg2006Nature,2007ApJ...657L..73G,2007ApJ...662.1111L,2009MNRAS.392...91V,2010tsra.confE.204R,2010MNRAS.406.1944W,2011A&A...525A..53G,Kovacevic2014}), for short bursts \citep[e.g.][]{Virgili2011,2015MNRAS.448.3026W}, and for both long and short bursts \citep[e.g.][]{2015ApJ...812...33S}. Additional properties that introduce further uncertainties are the instrumental sensitivity threshold, the field of view $\Omega_i$, and the operational time $T_i$ of $i$-detector.

Hereafter we neglect the possible redshift-evolution of the luminosity function. For $\Delta N_i$ events detected by various detectors in a finite logarithmic luminosity bin from $\log L$ to $\log L+\Delta\log L$, the total local event rate density between observed minimum, $L_{\rm min}$, and maximum, $L_{\rm max}$, luminosities, is \citep[e.g.][]{2015ApJ...812...33S}
\begin{equation}
\label{eq:rate}
\rho_0\simeq \sum_{i}\sum_{\log L_{\rm min}}^{\log L_{\rm max}}\frac{4\pi}{\Omega_i T_i}\frac{1}{\ln10}\frac{1}{g(L)}\frac{\Delta N_i}{\Delta\log L}\frac{\Delta L}{L}\ ,
\end{equation}
where
\begin{equation}
\label{eq:rate2}
g(L)=\int_{0}^{z_{\rm max}(L)}\frac{f(z)}{1+z}\frac{dV(z)}{dz}dz\ ,
\end{equation}
and the comoving volume is
\begin{equation}
\frac{dV(z)}{dz}=\frac{c}{H_0}\frac{4\pi d_L^2}{(1+z)^2[\Omega_M(1+z)^3+\Omega_{\Lambda}]^{1/2}}\ ,
\end{equation}
where $d_L$ is the luminosity distance. We set $f(z)=1$, namely we do not assume any redshift dependence of the GRB cosmic event rate density. The maximum volume within which the event of luminosity $L$ can be detected is defined by the maximum redshift $z_{\rm max}(L)$. The latter is computed, following \citet{Schaefer2007}, from the $1$~s-bolometric peak luminosity $L$, $k$-corrected from the observed detector energy band into the burst cosmological rest-frame energy band $1$--$10^4$~keV, and the corresponding $1$~s-threshold peak flux $f_{\rm th}$. This is the limiting peak flux for the burst detection \citep{Band2003}. With this, $z_{\rm max}$ can be defined from \citep[see, e.g.,][]{Zhang2,2014A&A...569A..39R}
\begin{equation}
f_{\rm th} = \frac{L}{4\pi d_{L}^{2} (z_{\rm max})k}.
\end{equation}

The possible evolution with the redshift of the GRB density rates have been analyzed in \citet{2016ApJ...832..136R} by separating the bursts into several redshift bins, following the method suggested in \citet{2015ApJ...812...33S}. In each redshift interval $z_j\leq z\leq z_{j+1}$, the integration limits of Eq.~(\ref{eq:rate2}) are replaced by $z_j$ and min$[z_{j+1}, z_{\rm max,j}(L)]$, where $z_{\rm max,j}(L)$ is the maximum redshift for the j$^{th}$ redshift bin. Finally, from Eq.~(\ref{eq:rate}) an event rate $\rho_{\rm 0}^{\rm z}$ in each redshift bin around $z$ is obtained.

We adopt the fields of view and operational times for the detectors: \emph{Beppo}-SAX, $\Omega_{\rm BS}=0.25$~sr, $T_{\rm BS}=7$~y; BATSE, $\Omega_{\rm B}=\pi$~sr, $T_{\rm B}=10$~y, HETE-2, $\Omega_{\rm H}=0.8$~sr, $T_{\rm H}=7$~y; \textit{Swift}-BAT, $\Omega_{\rm S}=1.33$~sr, $T_{\rm S}=10$~y; \textit{Fermi}-GBM, $\Omega_{\rm F}=9.6$~sr, $T_{\rm F}=7$~y. We adopt no beaming correction.



\end{document}